\documentclass{mn}
\usepackage{amsmath, amssymb, epsfig}
\begin{document}
\def\Mpc{{$\,h^{-1}\,{\rm Mpc}$}}
\def\Mpcinv {{$\,h^{3}\,{\rm Mpc}^{-3}$}}
\def\simlt{\mathrel{\rlap{\lower 3pt\hbox{$\sim$}} \raise
        2.0pt\hbox{$<$}}} \def\simgt{\mathrel{\rlap{\lower
        3pt\hbox{$\sim$}} \raise 2.0pt\hbox{$>$}}}
\def\simgt{\mathrel{\rlap{\lower 3pt\hbox{$\sim$}} \raise
        2.0pt\hbox{$>$}}} \def\simgt{\mathrel{\rlap{\lower
        3pt\hbox{$\sim$}} \raise 2.0pt\hbox{$>$}}}
\title[Cosmic Evolution of Quasar Clustering]
{Cosmic Evolution of Quasar Clustering: Implications for the Host Haloes}
\author[Cristiano Porciani, Manuela Magliocchetti, Peder Norberg]
{Cristiano Porciani$^{1}$,
Manuela Magliocchetti$^{2}$, Peder Norberg$^{1}$\\
$^{1}$ Institute for Astronomy, ETH H\"onggerberg, CH-8093 Z\"urich,
Switzerland\\
$^{2}$SISSA, Via Beirut 4, 34014, Trieste, Italy}
\maketitle
\vspace {7cm}

\begin{abstract}
We present detailed clustering measurements from the 2dF QSO Redshift Survey 
(2QZ) in the redshift range $0.8<z<2.1$.
Using a flux limited sample of $\sim 14,000$ objects with effective
redshift $z_{\rm eff}=1.47$, we estimate the quasar projected correlation
function for separations $1<r/{h^{-1} {\rm Mpc}}<20$.
We find
that the 2-point correlation function in real space 
is well approximated by a power law with slope $\gamma=1.5\pm 0.2$ and 
comoving correlation length $r_0=4.8^{+0.9}_{-1.5}$ \Mpc.
Splitting the sample into three subsets based on redshift, 
we find evidence for
an increase of the clustering amplitude with lookback time.
For a fixed $\gamma$, 
evolution of $r_0$ is detected at the $3.6\,\sigma$ confidence level.
The ratio between the quasar correlation function
and the mass autocorrelation function (derived adopting the concordance 
cosmological model) is found to be scale independent.
For a linear mass-clustering amplitude $\sigma_8=0.8$, 
the ``bias parameter'' decreases
from $b\simeq 3.9$ at $z_{\rm eff}=1.89$ to 
$b\simeq 1.8$ at $z_{\rm eff}=1.06$.
From the observed abundance and clustering, we 
infer how quasars populate dark-matter haloes of different masses.
We find that 2QZ
quasars sit in haloes with $M>10^{12} M_\odot$ and that
the characteristic mass of their host haloes is of the order of
$10^{13} M_\odot$.
The observed clustering is consistent with
assuming that the locally observed 
correlation between black-hole mass and host-galaxy circular
velocity is still valid at $z>1$.
From the fraction of haloes which contain active quasars, we infer that
the characteristic quasar lifetime is 
$t_{\rm Q}\sim {\rm a\ few}\times 10^{7}\, {\rm yr}$
at $z\sim 1$ and approaches $10^8$ yr at higher redshifts.
\end{abstract}


\section{Introduction}
Recent dynamical studies  have provided strong evidence for the existence of 
supermassive black holes in the centre of most nearby galaxies (for a review,
see e.g. Richstone et al. 1998).
The mass of the central black hole seems to correlate with the luminosity
and the velocity dispersion of the spheroidal stellar component 
(e.g. Magorrian et al. 1998; Gebhardt et al. 2000; Ferrarese \& Merritt 2000;
Tremaine et al. 2002).
The surprising tightness of the latter relation suggests the existence of a 
strong connection between the formation of supermassive black holes and
the assembly of galactic spheroids
(Silk \& Rees 1998; Haehnelt \& Kauffmann 2000; Kauffmann \& Haehnelt 2000;
Monaco, Salucci \& Danese 2000; Granato et al. 2004). 

The mounting evidence for the presence of supermassive black
holes in nearby galaxies supports the theoretical belief that 
quasars are powered by black-hole accretion
(Salpeter 1964; Zel'dovich \& Novikov 1964; Lynden-Bell 1969).
For instance,
the locally estimated mass density in black holes 
and the observed evolution of the quasar luminosity function
seem to be consistent with this hypothesis
(Haehnelt, Natarajan, Rees 1998; 
Fabian \& Iwasawa 1999; Salucci et al. 1999;
Yu \& Tremaine 2002; Wyithe \& Loeb 2003; 
Marconi et al. 2004). 
However, a detailed understanding of the physical processes leading to 
quasar activity (and their connection with galaxy formation) 
is still lacking. 
For this reason, even simple phenomenological models which are able
to reproduce the observational results by selecting which cosmic
structures could harbour quasars are of paramount importance. 

In the currently favoured cosmological model,
galaxies are expected to form within extended dark-matter haloes.
At every epoch,
the number density and clustering properties of the haloes
can be readily (and reliably) computed as a function of their mass.
It is therefore of great interest to try to establish a connection between 
these haloes and different classes of cosmic objects.
Even though
the distribution of light sources within haloes is determined
by complex physics, some of its properties can be computed
with a purely statistical approach.
For instance, 
one can use
the mean density and the clustering amplitude of a population of cosmic
objects to determine the lowest-order moments
of the ``halo occupation distribution function'', $P_N(M)$, which gives
the probability of finding a given number
of sources in a halo of mass $M$ (e.g. Scoccimarro et al. 2001 and
references therein).
A number of ``halo models'' have been presented in the literature. 
These have been 
successfully used to describe the adundance and clustering properties 
of galaxies at both low (Peacock \& Smith 2000; Seljak 2000; 
Scoccimarro et al. 2001; 
Marinoni \& Hudson 2002; Berlind \& Weinberg 2001; 
Yang, Mo \& van den Bosch 2003;
van den Bosch, Yang \& Mo 2003; Zehavi et al. 2003; Magliocchetti \& Porciani 
2003) and high (Bullock, Wechsler \& Somerville 2002; 
Moustakas \& Somerville 2002; Hamana et al. 2004; Zheng 2004) redshift. 
One of the main
goals of this paper is to use a similar approach to 
investigate how quasars populate dark matter haloes.
As previosly discussed, this requires an accurate
determination of the clustering properties of bright QSOs.
 
Since the first detection of quasar clustering,
(Shaver 1984; Shanks et al. 1987), 
a number of surveys (continuously improved in terms of homogeneity, 
completeness and size) have been used to measure the
2-point correlation function of bright QSOs
(Iovino \& Shaver 1988; Andreani \& Cristiani 1992; Mo \& Fang 1993; 
Shanks \& Boyle 1994; Andreani et al. 1994; Croom \& Shanks 1996;
La Franca, Andreani \& Cristiani 1998; Grazian et al. 2004).
The emerging picture is that quasars at $z\sim 1.5$ have a correlation
length $r_0\simeq 5-6$ \Mpc, similar to that of present-day galaxies.
It still is a matter of debate, however, whether $r_0$ significantly
evolves with redshift (Iovino \& Shaver 1988; Croom \& Shanks 1996; 
La Franca, Andreani \& Cristiani 1998). This uncertainty is due to
the joint effects of cosmic variance and small-number statistics:
given the sparseness of the quasar distribution, a typical sample 
includes from a few-hundred to a thousand objects. In consequence,
clustering is generally detected at a relatively low significance 
level ($3-4 \,\sigma$).

The development of efficient multi-object spectrographs has recently 
made possible a new generation of wide-area redshift surveys. 
Both the completed Two-degree Field (2dF) QSO Redshift Survey 
(Croom et al. 2004) and the on-going 
Sloan Digital Sky Survey (SDSS) Quasar Survey (Schneider et al. 2003)
list redshifts for tens of thousands of optically selected quasars.
A preliminary data release of the 2QZ has been used to
estimate the evolution and the luminosity dependence of the
quasar 2-point correlation function in redshift-space 
(Croom et al. 2001; 2002).
The final catalogue has been employed to measure the quasar power spectrum 
out to scales of $~500$ \Mpc\   (Outram et al. 2003; see also 
Hoyle et al. 2002) and to constrain the cosmological constant 
from redshift-space distortions (Outram et al. 2004).

In this paper, we study the clustering properties of
$\sim 14,000$ quasars extracted from the complete 2dF QSO Redshift Survey.
In particular, we compute the evolution of their
projected 2-point correlation function.
This quantity is not affected by the 
distortion of the clustering pattern induced by peculiar motions
as it measures the clustering strength as a function of quasar separation
in the perpendicular direction to the line of sight.
Using the largest quasar sample presently available, we 
are able to accurately
measure the real-space clustering amplitude in three redshift bins.
Our results reveal a statistically significant evolution 
of the clustering length with redshift.

We then combine our clustering study with the halo model 
to infer the mean number of optically selected quasars 
which are harboured by a virialized halo of given mass
(the halo occupation number)
and the characteristic quasar lifetime. 
%
Our results can be directly used to test
physical models for black-hole accretion. 

The layout of the paper is as follows.
In Section \ref{datasec} we present our quasar samples,
we measure their projected correlation function 
and estimate the corresponding bias parameters. 
In Section \ref{sechon},
we introduce the halo model and  
we discuss how the the halo occupation number, $N(M)$, is constrained 
by the observed abundance and clustering amplitude of 
optically bright quasars.  
Using the empirical correlation between black-hole mass and circular
velocity of the host galaxy (Ferrarese 2002),
in Section \ref{sol}, we present a new derivation
of the function $N(M)$ based on
the observed quasar luminosity function.
We then show that this model is in agreement with the clustering
measurements presented in Section \ref{datasec}.
In Section \ref{bayes}, we present a Bayesian study of the halo
occupation number which combines the results from Sections \ref{sechon} and 
\ref{sol}.
This allows us to fully constrain all the parameters of the halo model.
Estimates of the quasar lifetime are presented in Section \ref{lifetime}
while the evolution of the halo occupation number over the
redshift range $0.8<z<2.1$ is addressed in Section \ref{evo}.
Eventually, we discuss our results in Section \ref{discuss}
and conclude in Section \ref{summary}.

Throughout this work, we assume  
that the mass density parameter $\Omega_0=0.3$
(with a baryonic contribution $\Omega_{\rm b}=0.049$), 
the vacuum energy density parameter
$\Omega_\Lambda=0.7$ and the present-day value of the Hubble constant  
$H_0=100\,h\,{\rm km}\,{\rm s}^{-1}\,{\rm Mpc}^{-1}$ with $h=0.7$.
We also adopt a cold dark matter power spectrum with primordial spectral
index $n=1$  and 
$\sigma_8=0.8$ (with $\sigma_8$ the rms linear density fluctuation within a 
sphere with a radius of $8\, h^{-1}$ Mpc).
This is consistent with the most recent
joint-analyses of temperature anisotropies in the cosmic microwave background
and galaxy clustering
(see e.g. Tegmark et al. 2004 and references therein)
and with the observed quasar power spectrum (Outram et al. 2003).

\section{The Data}
\label{datasec}
In this section, we present the main properties of our dataset 
and compute its clustering properties as a function of redshift.

\begin{table*}
\begin{center}
\caption{Main properties of our datasets. The superscripts min, max and med
respectively denote the minimum, the maximum and the median value
of a variable.
\label{Tab:data}} 
\begin{tabular}{cccccccccc}
\hline
\hline
$z_{\rm min}$& $z_{\rm max}$ & $z_{\rm eff}$ &
$M^{\rm min}$ &
$M^{\rm max}$ & $M^{\rm med}$& $N_{\rm QSO}$ & $n_{\rm QSO}$ &
$b_{\rm J}^{\rm med}$\\
& & & \multicolumn{3}{c}{$M_{b_{\rm J}}-5\log_{10}h$} & &$10^{-6}\, h^3\, {\rm Mpc}^{-3}$ & & \\
\hline
$0.8$&$1.3$ & 1.06 & -25.32 & -21.72 &  -23.13 & 4928 & 
$8.54 \pm 0.47 \pm 0.85$
& 19.95\\ 
$1.3$&$1.7$& 1.51 & -25.97 & -22.80 &-23.84 & 4737 & $7.20\pm 0.35 \pm 0.72$
& 20.02\\
$1.7$&$2.1$& 1.89 & -26.44 & -23.37 & -24.30 & 4324 & $6.21\pm 0.26 \pm 0.62$
& 20.07\\
\hline
$0.8$&$2.1$& 1.47 & -26.44 & -21.72 & -23.82 & 13989 & $11.49\pm 1.52 \pm 1.15$
& 20.02\\
\hline
\end{tabular}
\end{center}
\end{table*}

\subsection{Quasar selection and sample definitions}
\label{sample}
The 2dF QSO Redshift Survey (2QZ) is a homogeneous database 
containing the spectra of 44,576 stellar-like objects
with $18.25 \le b_{\rm J}\le 20.85$ (Croom et al. 2004).
Selection of the quasar candidates is based on broadband colours 
($ub_{\rm J}r$) from APM measurements of UK Schmidt Telescope 
(UKST) photographic plates. Spectroscopic observations 
with the 2dF instrument (a multi-fibre spectrograph)
at the Anglo-Australian Telescope have been used to determine the intrinsic 
nature of the sources.
The full survey includes 23,338 quasars 
(the vast majority of which are endowed with a high-quality identification
and/or redshift) which span a wide redshift range ($0.3\simlt z \simlt 2.9$)
and are spread over 721.6 deg$^{2}$ on the sky
(see Croom et al. 2004 for further details).

The 2QZ is affected by incompleteness in a number of different ways
(for a detailed discussion see Croom et al. 2004 ).
In order to minimise systematic effects, 
we restrict our analysis to a sub-sample of
sources defined by a minimum spectroscopic sector completeness of 70
per cent.~\footnote{A sector is as a unique area on the sky 
defined by the intersection of a number of circular 2dF fields.}
Moreover, we
only consider regions of the 2QZ catalogue for which the photometric
completeness is greater than 90 per cent; this corresponds to a
redshift range $0.5<z<2.1$. Finally, we impose a cut in absolute
magnitude, so that we only consider quasars brighter than 
$M_{b_{\rm J}}-5\log_{10}h=-21.7$, which, assuming $h=0.7$,
is equivalent to $M_{b_{\rm J}}=-22.5$. Such an absolute magnitude
cut ensures the exclusion of quasars where the contribution from the
host galaxy may have led to a mis-identification of the source.

In order to detect possible evolutionary effects,
we want to subdivide our sample into three redshift bins.
In particular, we require that: (i) a similar number of quasars lies in each 
redshift bin, and (ii) each sub-sample covers a not too different interval
of cosmic time.
For this reason, we revise our initial sample selection by
imposing an additional redshift cut, so to keep only objects within
$0.8<z<2.1$. 
In fact,
the time interval covered by the redshift range $0.5<z<0.8$ (1.78 Gyr)
corresponds to 
one third of the total time elapsed between $z=2.1$ and $z=0.5$ (5.35 Gyr),
whereas the number of quasars with $0.5<z<0.8$ 
represents less than 10 per cent of the 
selected quasar sample. By restricting the analysis
to $0.8<z<2.1$, we can greatly improve on both the previously
mentioned conditions. Moreover, we obtain a sample for which the mean
number density of quasars is very weakly varying with redshift (as it can be
seen in Fig.~1 of Outram et al. 2003), since through this cut we remove
the largest mean number density variations as a function of redshift. The
drop in mean density is less than 60 per cent between $z=0.8$ and
$z=2.1$. 


With the above selection, we end up with nearly 14,000 quasars (split in two
regions -- the north galactic pole strip (NGP) and the
south galactic pole strip (SGP) -- with respectively $\sim7800$ and $\sim6100$ 
quasars each), of which 75 per cent reside in regions with total completeness
larger than 80 per cent.
In order to study the evolution of quasar clustering,
we divide this sample into three redshift slices.
To satisfy our previously mentioned criteria,
we end up choosing
the following three intervals: $0.8<z<1.3$, $1.3<z<1.7$ and
$1.7<z<2.1$, containing each between $\sim4300$ and $\sim4900$ quasars 
(see Table \ref{Tab:data}). 
We note that the time covered between $z=0.8$ and $z=1.3$ (1.91 Gyr) is nearly
twice the time covered between $z=1.3$ and $z=1.7$ (0.97 Gyr); this is however
unavoidable if we want to keep similar numbers of quasars in each
redshift interval. As the sample is magnitude limited, each redshift
interval will correspond to quasars of different intrinsic
luminosities, a point we address further in Section~\ref{sec:num_dens}.

\subsection{Number densities of quasars}
\label{sec:num_dens}

Croom et al. (2004) provide an analytical fit for the $b_{\rm J}$
quasar luminosity function, in the case of sources brighter than 
$M_{b_{\rm J}}-5\log_{10}h\ge-21.7$ and for $0.4<z<2.1$. 
They model the optical luminosity function as a
double power law in luminosity which, as a function of magnitude
(number of quasars per unit magnitude per unit volume), becomes
\begin{eqnarray}
\Phi(M_{b_{\rm J}},z)=\frac{\Phi^*}
{10^{0.4\,\beta_1
(M_{b_{\rm J}}-M^\star_{b_{\rm J}})}+{10^{0.4\,\beta_2
(M_{b_{\rm J}}-M^\star_{b_{\rm J}})}}},
\label{eq:lf}
\end{eqnarray}
where the evolution is encoded in the redshift 
dependence of the characteristic magnitude 
$M^\star_{\rm b_j}\equiv M^\star_{\rm b_j}
(z)=M^\star_{\rm b_j}(0)-1.08 \,k\,\tau(z)$
with $\tau(z)$ the fractional look-back time (in
units of the present age of the Universe) at redshift $z$
and $k$ a constant. 
A table with the best-fitting values for
the parameters $\beta_1$, $\beta_2$, $M^\star_{b_{\rm J}}(0)$ and $k$
is provided by Croom et al. (2004), together with their statistical
uncertainties.

Equation (\ref{eq:lf}) can be used to compute the
selection function of the 2QZ between $z_{\rm min} <z <z_{\rm max}$:
\begin{equation}
S(z,z_{\rm min},z_{\rm max})=\frac{\int^{M_{b_{\rm J}}^{\rm f}(z)}
_{M_{b_{\rm J}}^{\rm b}(z)}\Phi(M_{b_{\rm J}},z)\, {\rm d}M_{b_{\rm J}}}
{\int_{M_{b_{\rm J}}^{\rm min}}^{M_{b_{\rm J}}^{\rm max}}
\Phi(M_{b_{\rm J}},z)\, {\rm d}M_{b_{\rm J}}},
\label{eq:sz}
\end{equation}
where $M_{b_{\rm J}}^{\rm b}(z)$ and $M_{b_{\rm J}}^{\rm f}(z)$ denote, 
respectively, the 
brightest and faintest
absolute magnitudes which are detectable at redshift $z$. 
These are obtained by using the
K-correction from the composite quasar spectrum by Brotherton et al. (2001),
and by assuming fixed apparent magnitude limits: 
$b_{\rm J}^{\rm faint}= 18.25$ and $b_{\rm J}^{\rm bright}= 20.85$.
The integration limits in the denominator of equation (\ref{eq:sz}) are
$M_{b_{\rm J}}^{\rm min}={\rm min}_{(z_{\rm min},z_{\rm max})} 
M_{b_{\rm J}}^{\rm b}(z)$ and
$M_{b_{\rm J}}^{\rm max}={\rm max}_{(z_{\rm min},z_{\rm max})} 
M_{b_{\rm J}}^{\rm f}(z)$.
%
The comoving volume effectively surveyed is then given by
\begin{equation}
V_{\rm eff}(z_{\rm min},z_{\rm max})=
\Omega_{\rm s} \int_{z_{\rm min}}^{z_{\rm max}} 
    S(z,z_{\rm min},z_{\rm max}) 
\left|\frac{{\rm d}V}{{\rm d}z\,{\rm d}\Omega}\right|\,{\rm d}z\;,
\end{equation}
where $\Omega_{\rm s}$ denotes the solid angle covered by the survey
and $|{\rm d}V/{\rm d}z\,{\rm d}\Omega|$ is the Jacobian determinant of the
transformation between comoving and redshift-space coordinates, 
which gives the comoving volume element per unit redshift and solid angle.
We can then estimate the 
mean number density of quasars 
by writing
\begin{equation}
n_{\rm QSO}(z_{\rm min},z_{\rm max}) = \sum_{i=1}
^{N_{\rm QSO}} \frac{w_i}{V_{\rm eff}(z_{\rm min},z_{\rm max})}\;,
\label{eq:ng}
\end{equation}
where $N_{\rm QSO}$ and $\sum_{i} w_i$ are, respectively,
the total number of observed quasars 
in the range $z_{\rm min} <z <z_{\rm max}$
and its completeness weighted counterpart. 
Results obtained by applying equation (\ref{eq:ng}) are listed
in Table \ref{Tab:data}, where we summarize the main properties of our 
samples. 
%
Two types of errors are quoted for $n_{\rm QSO}$. Those listed first
are determined by independently varying the four parameters which define the
optical quasar luminosity function (i.e. $\beta_1$, $\beta_2$,
$M^\star_{b_{\rm J}}(0)$ and $k$) within their $1\,\sigma$ uncertainties
as reported by Croom et al. (2004).~\footnote{If the errors for the four parameters are statistically
independent, the quoted values for $\Delta n_{\rm QSO}$ approximately 
give $1\,\sigma$ uncertainties, 
whereas, if the parameters are correlated (which they most
certainly are), the quoted errors for $n_{\rm QSO}$ correspond to a higher
confidence interval.} On top of this error, we quote 
a $\sim 10$ per cent uncertainty on $n_{\rm QSO}$ due to the large-scale
distribution of quasars: as in Outram et al. (2003), we typically find a
$\sim 10$ per cent difference in the number counts between the SGP and NGP. 
As this
difference appears to be nearly constant as function of redshift, we
quote for all subsamples the same typical uncertainty due to large
scale structure. We note that our determination of $n_{\rm QSO}$ is 
independent of the normalisation constant $\Phi^\star$ appearing in 
equation (\ref{eq:lf}).

From the luminosity function we also compute the effective redshift of
the different samples as
$z_{\rm eff}= \sum_{i=1}^{N_{\rm QSO}} w_i\, z_i/\sum_{i=1}^{N_{\rm QSO}} 
w_i$ (see Table \ref{Tab:data}).

\subsection{Quasar projected correlation function}
\label{pcf}
The simplest statistic which can be used to quantify
clustering in the observed quasar distribution
is the 2-point correlation function in redshift space, 
$\xi^{\rm obs}(r_\perp,\pi)$.
This measures the excess probability over random to find a quasar pair
separated by $\pi$ along the line of sight
and by $r_\perp$ in the plane of the sky.
These separations are generally derived from the
redshift and the angular position of each source,
so that the inferred $\pi$ includes a contribution from peculiar
velocities. In consequence,
the reconstructed clustering pattern in comoving space comes out to 
be a distorted representation of the real one and 
$\xi^{\rm obs}(r_\perp,\pi)$ is found to be anisotropic.
To avoid this effect, and determine the quasar clustering amplitude in 
real space, one can then use the 
``projected correlation function''
which is obtained by integrating $\xi^{\rm obs}(r_\perp,\pi)$ 
in the $\pi$ direction:
\begin{equation} 
\frac{\Xi^{\rm obs}(r_\perp)}{r_\perp} =  
\frac{2}{r_\perp} \int_{0}^{\infty} 
\xi^{\rm obs}(r_\perp,\pi)\,{\rm d}\pi \,,
\label{proj}
\end{equation}
and it is therefore unaffected by redshift-space distortions. 
In this section, we measure this quantity for our quasar samples.

We start by building a catalogue of
unclustered points which has the same angular and
radial selection function as the data. The angular selection is trivially given
by the different completeness masks (see Croom et al. 2004 for further
details), and we modulate the number of random points 
laid down as a function of
spectroscopic and photometric completeness. The radial selection 
function is instead obtained by
heavily smoothing the observed quasar redshift distribution, ${\cal N}(z)$, or
even the observed quasar comoving distance distribution, ${\cal N}(r)$. 
Both uniform and Gaussian smoothings, with characteristic smoothing
length of several hundreds of \Mpc, have been used. The quoted
results are insensitive to the precise details of the modelling of the
radial selection function. This is partially due to the fact that the
volume covered by the quasar sample is very large and that there are
not many clear groups or clusters of quasars: the
quasar redshift distribution is rather smooth when compared to a
standard galaxy distribution (e.g. Fig.~1 of Outram et al. 2003 vs
Fig.~13 of Norberg et al. 2002). 
%

The quasar correlation function is then estimated by
comparing the probability distribution 
of quasar and random pairs
on a two-dimensional grid of separations $(r_\perp, \pi)$.
We use both 
the Landy-Szalay estimator (Landy \& Szalay 1993) and the Hamilton estimator
(Hamilton 1993):
\begin{equation}
\xi_{\rm{LS}}^{\rm obs}= \frac{ DD - 2DR + RR }{RR}\;,\ \ \ 
\xi_{\rm{H}}^{\rm obs} = \frac{DD\cdot RR}{(DR)^2}\,-\,1
\end{equation}
where $DD$, $DR$ and $RR$ are the suitably normalised numbers of  
weighted data-data, data-random and random-random pairs in 
each bin. 
\footnote{Note that, in this case, there is no need to use 
the standard $J_3$ (minimum variance) weighting scheme
(Efstathiou 1988) since the mean density of quasars,
$n_{\rm QSO}$, is so low that $1\,+\,4\,\pi\,J_3\,
n_{\rm QSO}\simeq 1$ for any reasonable quasar clustering
amplitude.}
As expected, the two estimators give comparable 
answers within the errors. For this reason, in what follows 
we only present results obtained with the Hamilton estimator.

With the current quasar sample, we find that a reliable measure of
$\xi^{\rm obs}(r_\perp,\pi)$ is only achievable on scales
$\pi\simlt 50$ \Mpc. 
In fact, the number of quasar pairs with small transverse
separations and large line-of-sight separation is very small. 
This is partially due to
the sparseness of the samples considered but also to the rareness
of large structures of quasars. 
Eventually, we compute the projected correlation function
using equation (\ref{proj}).
In order to avoid the measured signal to be dominated by noise, we
limit the integration to an upper limit, $\pi_{\rm max}$.
This limiting value needs to be sufficiently large
in order to give a reliable and meaningful measurement of
$\Xi^{\rm obs}(r_\perp)/r_\perp$ on the scales of interest 
(i.e. $r_\perp \simlt 20$ \Mpc) but also small enough to be less sensitive to 
noise. 
Using the redshift distortion models by Hoyle et al. (2002),
we find that $\pi_{\rm max}=45$ \Mpc\ fulfils both requirements.
All the data presented in this paper are obtained using this value.
In any case, we verify that our results are not sensitive to the precise value
adopted for $\pi_{\rm max}$.

\begin{figure*}
\centerline{\epsfig{figure=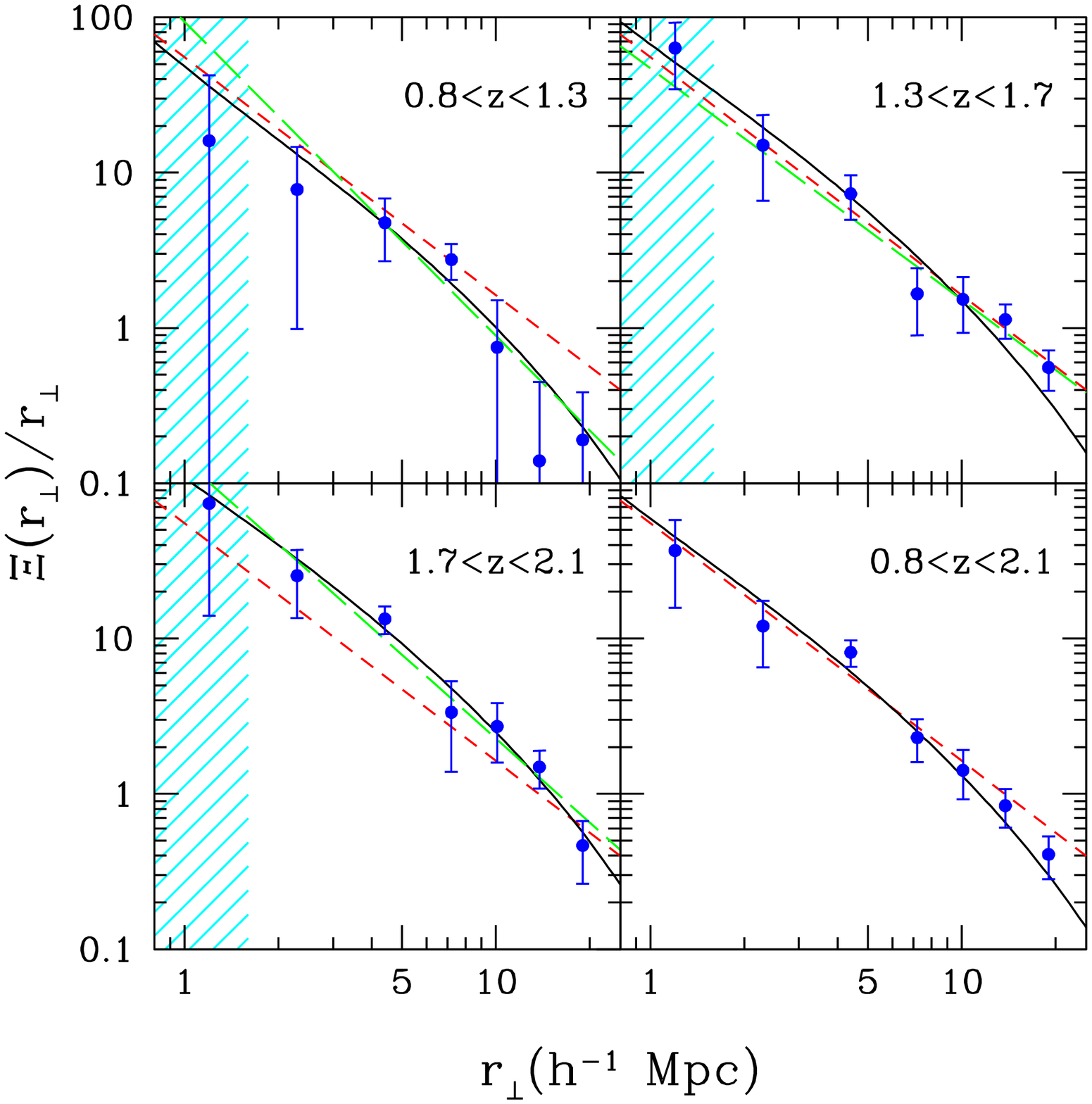,height=8cm}
\hspace{0.3cm}
\epsfig{figure=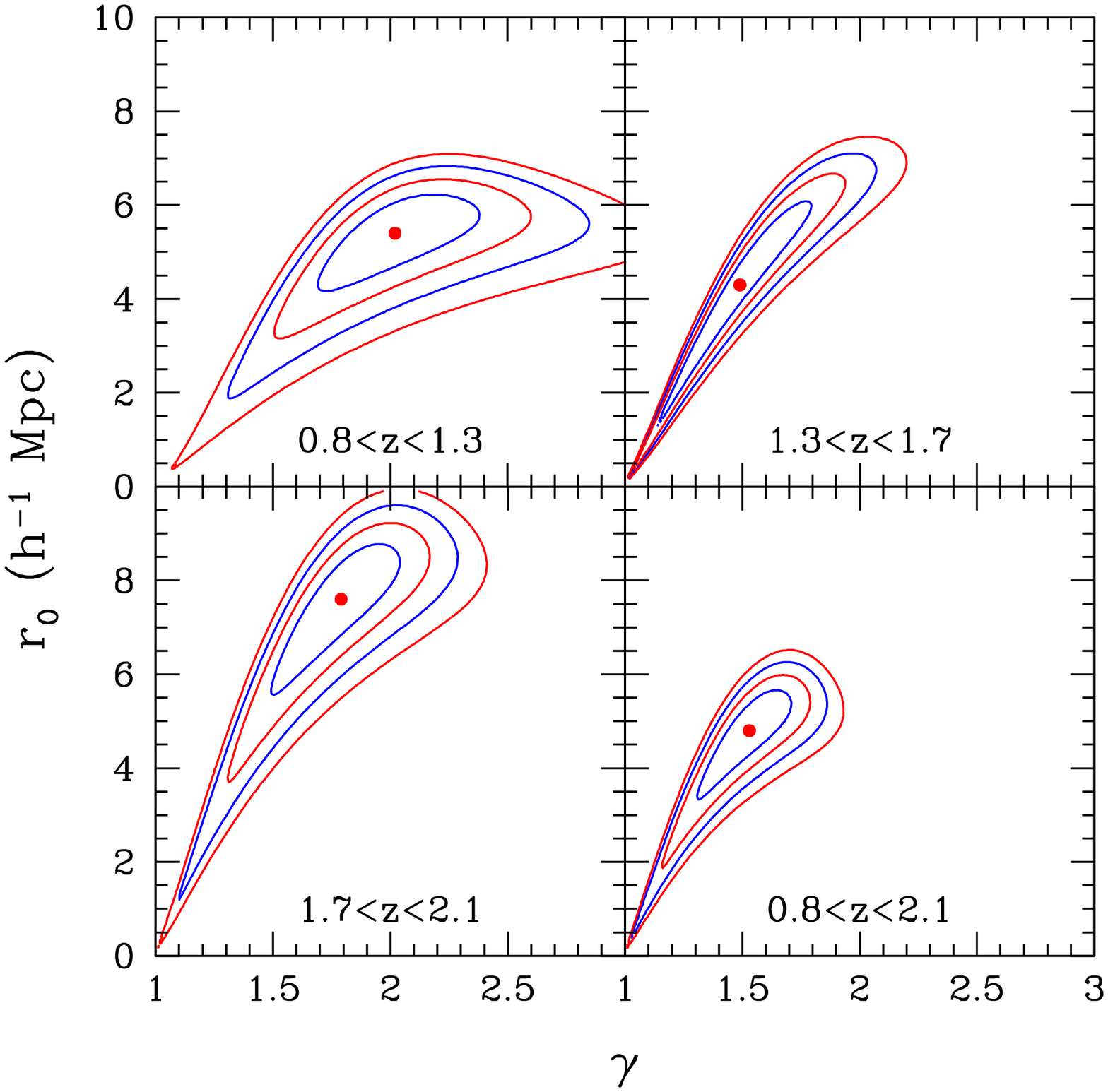,height=8cm}}
\caption{{\it Left panel:}
The projected correlation function for different samples
of quasars from the 2QZ (data with errorbars).
For each redshift sub-sample,
the best-fitting power law is represented with a long-dashed line.
For reference, 
the best-fitting power law for the full sample ($0.8<z<2.1$)
is shown with a short-dashed line.
The continuous lines represent the best-fitting constant-bias models
discussed in Section \ref{biassec}. These functions practically coincide
with the best-fitting halo-occupation models presented in 
Sections \ref{lss} and \ref{bayes}. 
Data in the shaded regions are derived from less than 20 quasar pairs and
are not considered for model fitting.
{\it Right panel:} Contour levels for the likelihood function obtained
by fitting the data with a power law model. 
The best-fitting models are marked with a dot and the lines
correspond to 4 different values of $\Delta \chi^2=\chi^2-\chi^2_{\rm min}$.
In particular, for Gaussian errors, the inner contours ($\Delta \chi^2=1$ and
2.3) mark the 68.3 per cent confidence levels for one and two parameters,
respectively. Similarly, the outer contours ($\Delta \chi^2=4$ and
6.17) correspond to the 95.4 per cent confidence levels for one and two
parameters.
\label{Fig:data}}
\end{figure*}

\subsection{Error estimates for clustering measurements} 

It is common practice
to estimate errors on the clustering measurements
from mock surveys based on galaxy formation models.
However, for quasars, such catalogues are not publicly available.
We therefore opt for a bootstrap resampling technique.
For each redshift-interval we divide both the NGP and the SGP samples into 
eight equal-volume regions, and 
we measure the clustering properties of each sub-sample.
For ease of calculation, the division is just based
on redshift and is such that the depth of each
region is larger than the adopted value for $\pi_{\rm max}$.
Since the number density is roughly constant as function of
redshift, each of these regions approximately contains the same
number of quasars. 
We then build 1000 bootstrap-samples, each of them composed by sixteen 
sub-samples (eight for each strip)
randomly drawn (allowing repetitions) from the
set described above.
We measure the projected correlation function for each artificial sample
by appropriately averaging over the number of quasar and random pairs 
of the sub-samples
(and not over individual quasar clustering measurements). 
For each $r_\perp$, we identify
the rms variation of $\Xi$ over the bootstrap-samples
with the $1\,\sigma$ error for the projected correlation function.

Our method for estimating errors relies on the fact that our dataset 
is statistically representative of the quasar distribution in the Universe.
However, this cannot be true for bins of spatial separations which
contain just a few quasar pairs. 
Therefore, in what follows, we ignore clustering results obtained by less
than 20 quasar pairs. Depending on the sample, this corresponds to 
$r_\perp<1-2$ \Mpc. Note that, on such
scales, corrections for close pair incompleteness
due to fibre collisions (Croom et al. 2004) should be also
taken into account.

\begin{table*}
\begin{center}
\caption{Best-fitting power law and constant-bias models for the four
quasar samples. The goodness of each fit is measured by the quantity
$\chi^2_{\rm min}/{\rm dof}$ which gives the minimum value assumed by
the chi-square statistic divided by the number of degrees of freedom.
The symbol $M_b$ denotes the halo mass which matches the observed clustering 
amplitude. 
\label{Tab:PL}} 
\begin{tabular}{ccccccccccccc}
\hline
\hline
$z_{\rm min}$&$z_{\rm max}$ & $r_0$ & $\gamma$ & $
\displaystyle{\frac{\chi^2_{\rm min}}{{\rm dof}}}$ &
$r_0^{(\gamma=1.53)}$ & $\displaystyle{\frac{\chi^2_{\rm min}}{{\rm dof}}}$ &
$r_0^{(\gamma=1.8)}$ & 
$\displaystyle{\frac{\chi^2_{\rm min}}{{\rm dof}}}$
& $b$& $\displaystyle{\frac{\chi^2_{\rm min}}{{\rm dof}}}$ & $\log_{10}\displaystyle{\frac{M_b}{M_\odot}}$\\
& &$(h^{-1}\, {\rm Mpc})$ & & & $(h^{-1}\, {\rm Mpc})$ & &
$(h^{-1}\, {\rm Mpc})$ & & \\\hline
$0.8$&$1.3$ & $5.4^{+0.9}_{-1.3}$ & $2.02^{+0.36}_{-0.33}$ & 2.36/2 & 
$3.4^{+0.6}_{-0.7}$ & 4.52/3 & 
$4.7^{+0.7}_{-0.7}$ & 2.85/3 & $1.80^{+0.20}_{-0.24}$& $1.96/3$ & 
$12.80^{+0.20}_{-0.33}$
\vspace*{0.1cm}\\ 
$1.3$&$1.7$ & $4.3^{+1.8}_{-2.0}$ & 
$1.49^{+0.32}_{-0.35}$ & 1.25/2 & $4.6^{+0.4}_{-0.5}$ & 1.27/3 & 
$6.0^{+0.4}_{-0.5}$ & 2.20/3 & $2.62^{+0.18}_{-0.19}$& $3.18/3$ &
$13.00^{+0.11}_{-0.12}$
\vspace*{0.1cm}\\
$1.7$&$2.1$ & $7.6^{+1.2}_{-2.1}$ & 
$1.79^{+0.25}_{-0.29}$ & 2.04/2 & 
$5.9^{+0.7}_{-0.7}$ & 2.85/3 & 
$7.6^{+0.8}_{-0.8}$ & 2.04/3 & $3.86^{+0.32}_{-0.35}$& $0.84/3$ &
$13.26^{+0.11}_{-0.14}$
\vspace*{0.1cm}\\
\hline 
$0.8$&$2.1$ & $4.8^{+0.9}_{-1.5}$ & 
$1.53^{+0.19}_{-0.21}$ & 0.13/2 & $4.8^{+0.6}_{-0.6}$ & 
0.13/3 & 
$5.4^{+0.5}_{-0.5}$ & 2.54/3 &$2.42^{+0.20}_{-0.21}$& $0.57/3$ &
$12.91^{+0.13}_{-0.16}$
\vspace*{0.1cm}\\ 
\hline
\end{tabular}
\end{center}
\end{table*}

\subsection{Results}
\label{results}
The projected correlation function we obtained for the different redshift bins
(and for the total sample) is presented in the left panel of  
Fig. \ref{Fig:data}. A clear evolutionary trend emerges from the data:
the clustering amplitude for the high-redshift sample is nearly a factor
of 2 (3) higher than for the total sample (low-redshift one).

As a zero-th order approximation, we fit our results with a power-law
functional form. This phenomenological description has been commonly
used in the literature and allows us to compare our results with previous 
studies. More detailed modelling is presented in the next sections.
Here, we assume that the quasar 2-point correlation function 
scales as
\begin{equation}
\xi(r)=\left(\frac{r_0}{r} \right)^\gamma
\end{equation}
where $r$ denotes the comoving separation between quasar pairs.
The corresponding projected correlation function is obtained through
the simple integral relation,
\begin{equation}
\Xi(r_\perp)=2\int_{r_\perp}^{\infty}\frac{r\,\xi(r) }{(r^2-r_\perp^2)
^{1/2}}\,{\rm d}r\;,
\label{eq:xibar}
\end{equation}
which, in the power-law case, reduces to
\begin{equation}
\frac{\Xi(r_\perp)}{r_\perp}=
\frac{\Gamma(1/2)\,\Gamma[(\gamma-1)/2]}{\Gamma(\gamma/2)}
\,\left(\frac{r_0}{r_\perp} \right)^\gamma
\end{equation}
with $\Gamma(x)$ the Euler's Gamma function.
We use a minimum least-squares method 
(corresponding to a maximum likelihood method in the case of Gaussian errors)
to determine the 
values of $r_0$ and $\gamma$ that best describe the clustering data. 
A principal component analysis (see e.g. Porciani \& Giavalisco 2002) 
is used here to deal with correlated errorbars. 
The principal components of the errors
have been computed by diagonalizing the covariance matrix obtained by 
resampling the data with the bootstrap method described in the previous 
section.
The objective function (the usual $\chi^2$ statistic) 
has been obtained by considering the first four 
principal components which, for each redshift interval, account for more
than 85 per cent of the variance.

The best-fitting values for $\gamma$ and $r_0$ are given in Table 
\ref{Tab:PL} and the corresponding projected correlation functions are 
overplotted to the data in the left panel of Fig. \ref{Fig:data}.
Contour plots of the $\chi^2$ functions
are shown in the right panel of Fig. \ref{Fig:data}. Note that
the correlation length, $r_0$, and the slope of the correlation
function, $\gamma$, are strongly covariant: in order to accurately describe
the data, larger values of $r_0$ need to be associated with steeper
slopes.
The best-fitting slope for the whole quasar sample, 
$\gamma=1.53^{+0.19}_{-0.21}$, is in very good agreement
with the redshift-space analysis by Croom et al. (2001) who found 
$\gamma=1.56^{+0.10}_{-0.09}$ at a mean redshift of $\bar{z}=1.49$.
This is not surprising,
since we only consider scales that are in the quasi-linear and linear
regime of gravitational instability where the correlation function
in real and redshift-space are proportional to each other
(Kaiser 1987).
Accordingly, the comoving correlation length we find in real space,
$r_0=4.8^{+0.9}_{-1.5}$~\Mpc, is, as expected, slightly smaller than
its redshift-space counterpart, $5.69^{+0.42}_{-0.50}$~\Mpc.~\footnote{Croom et al. (2001) assume statistically independent Poisson 
errorbars for the correlation function at different separations. This explains
the factor of 2 between their and our estimate of the uncertainty for 
$r_0$ and $\gamma$.} 

As previously discussed,
visual inspection of Fig. \ref{Fig:data} suggests that the quasar
clustering amplitude at $1.7<z<2.1$ is nearly a factor of 2 
higher than that obtained for the whole sample.
Two sigma evidence for an increase in the clustering
amplitude of optically selected quasar between $z\sim 1$ and $z\sim 1.8$
has been presented by La Franca, Andreani \& Cristiani (1998).  
However, given their sample size
(a few hundred quasars) it is not clear whether
the detected evolution is real or it is spuriously created by
cosmic variance effects (see e.g. the discussion in Croom et al. 2001 who,
using a preliminary data release of the 2QZ, 
found a that the redshift-space clustering amplitude at $z=2.7$ 
is a factor 1.4 higher than at $z=0.7$ which is marginally significant).
It is therefore interesting to try to quantify the evolution of the 
clustering amplitude in our large quasar sample. 
In order to facilitate the comparison among the different subsamples
(and with previous studies), we report in Table \ref{Tab:PL} 
the 68.3 per cent confidence intervals for $r_0$ obtained
by assuming $\gamma=1.53$ ($r_0^{(\gamma=1.53)}$) and $\gamma=1.80$
($r_0^{(\gamma=1.80)}$). 
When fixing the slope, we note a steady increase of the quasar correlation
length with $z$.
Within this approximation, clustering evolution with 
redshift is detected at the $\sim 3.6\, \sigma$ confidence level.
However, since we are dealing with a flux-limited sample
(where the highest-redshift objects have on average the highest
intrinsic luminosities), it is not possible to 
say whether this evolution of $r_0$ 
corresponds to a real change in the quasar population or
to clustering segregation with luminosity.
By analysing quasar clustering in the range $0.3<z<2.9$ as a function of 
apparent 
luminosity in the preliminary data release catalogue of the 2QZ,
Croom et al. (2002) found weak ($\simeq 1\,\sigma$) evidence 
for the brightest third of quasars on the sky to be more clustered than 
the full data set.
Even though the different selection criteria prevent a 
direct comparison, we find a statistically more significant change in 
the clustering strength among our sub-samples than that reported in Croom et 
al. (2002). 
We defer a detailed analysis of the luminosity dependence of
quasar clustering to future work.
\begin{figure*}
\centerline{\epsfig{figure=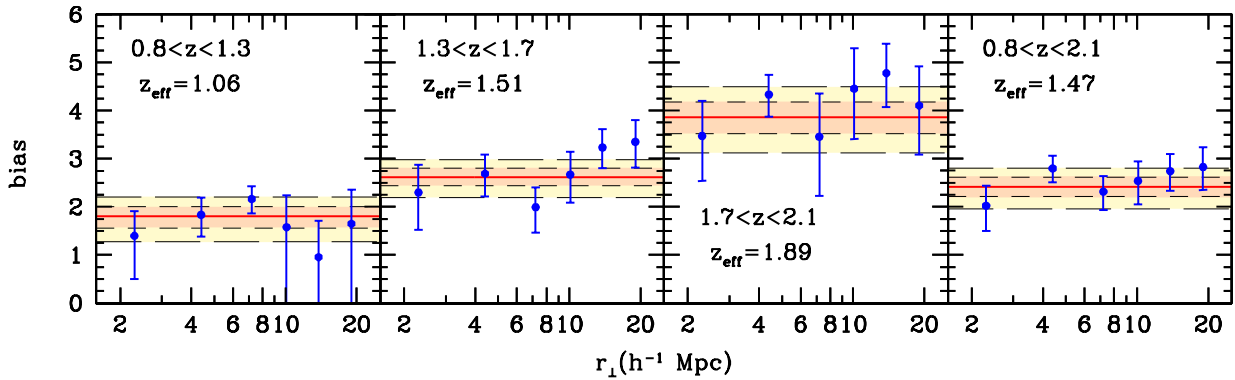,width=18cm}}
\caption{Quasar-to-mass bias function derived by
applying equation (\ref{biasdef}) 
to quasar
samples within different redshift ranges (points with errorbars).
The continuous line shows the best fitting constant value.
Dashed lines indicate the values of the bias for which 
$\Delta\chi^2=1$ (short dashed) and 
$\Delta\chi^2=4$ (long dashed).
\label{Fig:bias}}
\end{figure*}

\subsection{QSO vs galaxy clustering}
\label{qsovsgal}
How do our results on the spatial distribution of quasars compare with 
galaxy clustering at similar redshifts?
Until very recently, only rather small
samples of high-redshift galaxies were available 
and any attempt to determine their clustering properties was  
most probably hampered by cosmic variance 
(e.g. Le F\`evre et al. 1996; Carlberg et al. 1997; Arnouts et al. 1999;
Magliocchetti \& Maddox 1999).
The advent of 
color-selected surveys has allowed the detection of rich and homogeneous 
samples of high-redshift galaxies over relatively large volumes.
We want to compare the results obtained from the largest samples 
presently available with those obtained from our sample of quasars.
A number of studies have shown that 
Lyman-break galaxies at $z\sim 3$ are strongly clustered 
(e.g. Porciani \& Giavalisco 2002 and references therein).
Both their correlation length, $r_0\sim 4$~\Mpc, and
the slope of $\xi$, $\gamma\sim 1.5$, 
(Porciani \& Giavalisco 2002; Adelberger et al. 2003) 
are remarkably similar to 
the values obtained from our quasar sample.
On the other hand, star-forming galaxies at $z\simeq 1$ 
(detected by exploiting the Balmer break in their spectra)
are found to be slightly less clustered: $r_0\sim 3$~\Mpc \
with $\gamma\sim 1.7$ (Adelberger 2000).
Similarly,
the galaxy-galaxy correlation function from the 
DEEP2 Galaxy Redshift Survey  at $z_{\rm eff}=1.14$
is well described by a power law with $\gamma=1.66\pm 0.12$ and
$r_0=3.1\pm 0.7$~\Mpc \ (Coil et al. 2004). 
Evidence that early-type galaxies 
are more clustered than late-type ones (Firth et al. 2002; 
Coil et al. 2004) might
help reconciling these results with those inferred from the quasar population.
Extremely red galaxies at $z\sim 1$ have been found to be exceptionally 
strongly clustered.
Assuming $\gamma=1.8$ (as inferred from their angular clustering) one finds
$r_0 \sim 12$ \Mpc\  (Daddi et al. 2001; Firth et al. 2002; Roche, Dunlop \& 
Almaini 2003). 

\subsection{QSO vs dark-matter clustering}
\label{biassec}
To study how the spatial distribution of quasars relate
to the underlying matter distribution, 
we introduce the quasar-to-mass bias function:
\begin{equation}
b(r_\perp,z_{\rm eff})=\left[
\frac{\Xi(r_\perp,z_{\rm min}<z< z_{\rm max})}
{\Xi_{\rm m}(r_\perp,z_{\rm eff})}\right]^{1/2}
\label{biasdef}
\end{equation}
where $\Xi_{\rm m}$ is the projected correlation function of the mass 
distribution in the assumed cosmology computed as in Peacock \& Dodds (1996).
Fig. \ref{Fig:bias} shows our results for the different redshift bins.
Errorbars at different spatial separations are not statistically independent.
As previously described, we fit the data with a constant function
by using the principal components of the bootstrap errors 
(which shows that datapoints  at $r_\perp>10$ \Mpc\ are strongly correlated).
Results and the corresponding $1\,\sigma$ uncertainties 
are listed in Table \ref{Tab:PL}.
We find that $b$ steadily increases with $z$. This statistically
significant trend is mostly due to the rapid evolution
of the mass autocorrelation function. 

It is interesting to determine the halo mass, $M_{\rm b}$, which corresponds
to the observed quasar clustering amplitude (i.e. as if
all quasars would reside in haloes with a fixed mass).
We find that, for all the subsamples, 
 $M_{\rm b}$ is of the order of $10^{13} M_\odot$ and that it
slightly increases with $z$ (see Table \ref{Tab:PL}).

\section{The Halo Occupation Number of Quasars}
\label{sechon}
In this section, we present the halo model for the spatial distribution
of quasars.
After introducing the basic notation, we describe the main features
which characterize our model.~\footnote{
Further details can be found in Magliocchetti \& Porciani (2003) where
we used a similar tool to study the 
clustering properties of galaxies with different spectral types
in the 2dF Galaxy Redshift Survey (2dFGRS).}
We then use the number density and the projected correlation function
determined in Section \ref{datasec} 
to constrain the free parameters of the halo model. This 
allows us to determine the way quasars populate 
dark-matter haloes of different masses.

\subsection{The halo model}
\label{halomodel}
It is generally believed that
quasars are powered by mass accretion onto supermassive black holes
lying at the centre of galaxies.
Cold dark matter (CDM) models for structure formation 
predict that galaxies form within extended dark-matter haloes.  
The number density and clustering properties of these haloes 
can be easily computed, at any redshift, 
by means of a set of analytical tools which have been tested and 
calibrated against numerical simulations (e.g. Mo \& White 1996; 
Sheth \& Tormen 1999). 
In consequence, the problem of discussing the
abundance and spatial distribution of quasars can be reduced to studying how 
they populate their host haloes.
The basic quantity here is the halo occupation distribution function, $P_N(M)$,
which gives the probability of finding $N$ quasars within a single halo
as a function of the halo mass, $M$. 
Given the halo mass function $n(M)$
(number of dark-matter haloes per unit mass and volume), the mean 
value of the halo occupation
distribution $N(M)=\langle N \rangle(M)=\sum_N N\, P_N(M)$ 
(which from now on will be referred to as the halo occupation number) 
completely 
determines the mean comoving number density of quasars:
\begin{equation}
\bar{n}=\int n(M)\,N(M)\,{\rm d}M\;.
\label{eq:avern}
\end{equation}
Analogous relations, involving higher-order moments of $P_N(M)$,
can be used to derive the clustering properties of quasars in the halo 
model. For instance,
the 2-point correlation function can be written as 
the sum of two terms
\begin{equation}
\label{eq:xi}
\xi(r)=\xi^{1 \rm h}(r)+\xi^{2 \rm h}(r)\;.
\end{equation}
The function $\xi^{1\rm h}$, 
which accounts for pairs of quasars residing within the same halo, 
depends on the second factorial moment of the halo occupation distribution,
$\Sigma^2(M)=\langle N(N-1) \rangle(M)$ and on the spatial distribution
of quasars within their host haloes, $\rho({\bf x};M)$,~\footnote{
This is normalised in such a way that 
$\int_0^{r_{\rm vir}} \rho({\bf y};M)d^3y=1$
where $r_{\rm vir}$ is the virial radius which is assumed to mark the 
outer boundary of the halo.}
\begin{eqnarray}
\xi^{1\rm h}(r)=
\int \frac{n(M)\,\Sigma^2(M)}{\bar{n}^2}\,{\rm d}M
\int \rho({\bf x};M)\,\rho({\bf x}+{\bf r};M)\,{\rm d}^3x
\;.
\label{eq:xi1h}
\end{eqnarray}
On the other hand, the 
contribution to the correlation
coming from quasars in different haloes,  $\xi_{\rm QSO}^{2\rm h}$,
depends on $N(M)$ and $\rho({\bf x};M)$ as follows
\begin{eqnarray}
\xi^{2\rm h}&&\!\!\!\!\!\!\!\!\!\!\!\!\!(r)=
\int \frac{n(M_1)\,N(M_1)}{\bar{n}}
\,{\rm d}M_1 \int \frac{n(M_2)\,N(M_2)}{{\bar{n}}}\,{\rm d}M_2 \nonumber \\
&\!\!\!\!\!\!\times&\!\!\!\!\! \int
\rho({\bf x}_1;M_1)\,
\rho({\bf x}_2;M_2)\, 
\xi_{\rm h}({\bf r}_{12}; M_1, M_2) \,{\rm d}^3r_1\,{\rm d}^3 r_2
\;,  
\label{eq:xi2h}
\end{eqnarray}
where 
${\bf r}_i$ marks the position of the centre of each halo,
${\bf r}_{12}={\bf r}_2-{\bf r}_1$ is the separation between the haloes,
${\bf x}_i$ denotes the quasar position with respect to each halo centre
(hence ${\bf r}={\bf r}_{12}+{\bf x}_2-{\bf x}_1$), 
and $\xi_{\rm h}(r_{12}; M_1, M_2)$ is 
the cross-correlation function of haloes of mass $M_1$ and $M_2$, separated
by $r_{12}$.
For separations which are larger than the virial radius of the typical
quasar-host halo, the 2-halo term dominates the correlation function.
In this regime, $\xi_{\rm h}(r; M_1, M_2)$ scales proportionally to
the mass autocorrelation function, $\xi_{\rm m}(r)$, as 
$\xi_{\rm h}(r; M_1, M_2)\simeq 
b(M_1)\,b(M_2)\,\xi_{\rm m}(r)$,
with $b(M)$ the linear-bias factor of haloes of mass $M$
(Cole \& Kaiser 1989; Mo \& White 1996; Catelan et al. 1997;
Porciani et al. 1998). 
As a consequence of this, the large-scale behaviour of the 
quasar correlation function also comes out to be $\xi(r)\simeq 
b_{\rm eff}^2\,\xi_{\rm m}(r)$, with
\begin{eqnarray}
b_{\rm eff}=\frac{\int b(M)\,N(M)\,n(M)\,{\rm d}M}
{\int N(M)\, n(M)\,{\rm d}M}\;.
\label{beff}
\end{eqnarray}

Note that all the different quantities introduced in this section depend on 
the redshift $z$, 
even though we have not made it explicit in the equations.

In order to use the halo model to study quasar clustering, 
one has to specify a number of functions describing
the statistical properties of the population of dark-matter haloes.
In general, these have either been obtained analytically and then calibrated
against N-body simulations, or directly extracted from numerical experiments.
For the mass function and the linear bias factor of dark-matter haloes
we adopt here the model by Sheth \& Tormen (1999).
We approximate 
the 2-point correlation function of dark-matter haloes
with the function
(see e.g. Porciani \& Giavalisco 2002; Magliocchetti \& Porciani 2003)
\begin{equation}
\xi_{\rm h}(r;M_1,M_2)=
\begin{cases}
b(M_1)\,b(M_2)\,\xi_{\rm m}(r)&\text{if $r\ge r_1+r_2$} \\ 
-1 &\text{if $r<r_1+r_2$} 
\end{cases}
\label{eq:xihalo}
\end{equation}
where the mass autocorrelation function, $\xi_{\rm m}(r)$,
is computed using the method introduced by Peacock \& Dodds (1996) 
which, for our purposes, 
is sufficiently accurate both in the linear and non-linear regimes.~\footnote{In principle, 
for separations where $\xi_{\rm m}(r)\sim 1$, non-linear terms should
be added to equation (\ref{eq:xihalo}) 
(Mo \& White 1996; Catelan et al. 1997; Porciani et al. 1998). 
However, for the haloes and redshifts of interest,
these can be safely neglected.}
For small separations, 
equation (\ref{eq:xihalo}) accounts for spatial exclusion between haloes
(e.g. 2 haloes cannot occupy the same volume).
In Section \ref{ssc}, where we discuss the small-scale clustering of quasars,
we identify the Eulerian zone of exclusion of a given 
halo, $r_i$, with its virial radius.
At the same time, we assume that 
quasars trace the dark-matter distribution
and adopt, for $\rho({\bf x};M)$ a Navarro, Frenk \& White (1997; NFW) profile 
with concentration parameter obtained from equations (9) and 
(13) of Bullock et al. (2001).
Note that, in all the other sections of this paper,
we only consider the large scale distribution of 
quasars ($r\simgt 2$ \Mpc), where exclusion effects and 
density profiles do not affect the predictions of the halo model for $\xi$.

\subsection{Clustering on the light-cone} 
Equations (\ref{eq:xi}), (\ref{eq:xi1h}) and (\ref{eq:xi2h}) describe
the clustering properties of a population of cosmic objects selected
in a three-dimensional spatial section taken at constant
cosmic time in the synchronous gauge. However, deep surveys like the 2QZ
span a wide interval of lookback times and 
the equations we have introduced above do not apply in this case.

A number of authors have discussed 2-point statistics of objects lying 
on the light-cone of the observer 
(e.g. Matarrese et al. 1997, Yamamoto \& Suto 1999, 
Moscardini et al. 2000 and references therein). These works have shown that
the observed correlation function can be written as the weighted average: 
\begin{equation}
\xi^{\rm obs}(r)=\frac{\int_{z_{\rm min}}^{z_{\rm max}}
{\cal W}(z)\, \xi(r,z)\,{\rm d}z}{\int_{z_{\rm min}}^{z_{\rm max}}
{\cal W}(z)\,{\rm d}z}\;,
\label{lc}
\end{equation}
where ${\cal W}(z)={\cal N}(z)^2 ({\rm d}V/{\rm d}z)^{-1}$, 
with ${\cal N}(z)$ the actual
redshift distribution of the objects in the sample and 
${\rm d}V/{\rm d}z$ the Jacobian between comoving volume
and redshift. Note that equation (\ref{lc}) only holds for scales $r$ over 
which: {\it i)} ${\cal N}$ is nearly constant and 
{\it ii)} $\xi$ does not significantly evolve over the time $r/[(1+z)c]$
(where $c$ denotes the speed of light in vacuum).
Within the range of separations covered by our dataset, 
both the conditions are satisfied for our quasar sample.

Combining equations (\ref{eq:xibar}), (\ref{eq:xi}) and (\ref{lc}),  
we compute $\Xi^{\rm obs}(r_\perp)$ in our 4 redshift intervals
for a large set of $N(M)$ models and then compare the results with 
$\Xi(r_\perp,z_{\rm eff})$, the constant-time 
correlation function 
evaluated by using equations (\ref{eq:xibar}) and (\ref{eq:xi}) 
at the effective redshift of each sub-sample. 
In all cases we find extremely good agreement between the two functions.
Even for the widest redshift bin, $0.8<z<2.1$, we find a
maximum discrepancy of 2 per cent which is negligibly small with respect
to the typical error associated with the observed correlation function.
Therefore, in what follows, 
we use $\Xi(r_\perp,z_{\rm eff})$ to compare the predictions of different
models with the clustering data. 
This greatly simplifies (and speeds-up) the model fitting procedure.
As an additional test,
the comparison between $\Xi^{\rm obs}(r_\perp)$ and 
$\Xi(r_\perp,z_{\rm eff})$ is repeated for all the best-fitting models
that are discussed in the forthcoming sections and no significant difference
is found.

\subsection{The halo occupation number}
\label{sshon}
The final, key ingredient needed to describe the clustering properties of 
quasars is their halo occupation distribution function. 
In the most general case, $P_N(M)$ is entirely specified by  
all its moments which, in principle, could be observationally determined
by studying quasar clustering at any order. 
Regrettably, as we have already shown in Section \ref{results}, 
quasars are so rare that
their 2-point function is already very poorly determined so that it is not
possible to accurately measure higher-order statistics.
As in Magliocchetti \& Porciani (2003), we overcome this problem by assuming 
a predefined functional form for the lowest-order moments of $P_N(M)$.
It is, in fact, convenient to describe the halo occupation number, $N(M)$, 
and (if necessary) its associated scatter, $\Sigma^2(M)$, in terms
of a few parameters whose values will then be constrained by the data.

\label{phen}
%
%
We consider here the ``censored'' power-law model,
\begin{equation}
N(M)=N_0 \left(\frac{M}{M_0}\right)^\alpha\, \Theta(M-M_0)
\label{oldhon}
\end{equation}
(where $\Theta(x)$ 
denotes the Heavyside probability distribution function)
which has been widely
used in the literature to describe galaxy clustering 
(e.g. Magliocchetti \& Porciani 2003 and references therein).
In this case,
the halo occupation number vanishes for $M<M_0$ and scales as a power law
of the halo mass for $M>M_0$. The parameter $N_0$ gives the mean number of 
objects contained in a halo of mass $M_0$.~\footnote{Note that equation (\ref{oldhon}) is more general than the commonly
used $N(M)=(M/M_1)^\alpha\, \Theta(M-M_0)$ which, for $\alpha=0$,
automatically implies $N(M)=1$ at any $M>M_{0}$.}
Studies of the local galaxy population with
both hydrodynamical simulations and semianalytic models for galaxy formation 
are consistent with equation  (\ref{oldhon}) 
(e.g. Sheth \& Diaferio 2001; 
Berlind \& Weinberg 2002; Berlind et al. 2003).
We assume that the same parameterization is adequate for quasars at
$z\simgt 1$.
Given the observational evidence for a correlation 
between black-hole and host-halo masses 
(Ferrarese 2002; see also the discussion in Section \ref{halomass}), 
it is reasonable to expect the presence of a 
threshold mass for the host haloes of a quasar population with a given
minimum luminosity.
At the same time, a power-law 
scaling with an unspecified index $\alpha$ for $M>M_0$ is general enough
to explore a wide range of possibilities. 
%
%

It would be ideal to test equation (\ref{oldhon}) against physical
models for quasar activity.
A number of authors recently developed simplified schemes to
include black-hole accretion in galaxy formation models
(e.g. Kauffmann \& Haehnelt 2002; Enoki, Nagashima \& Gouda 2003;
Menci et al. 2003; Di Matteo et al. 2003 and references therein).
Regrettably, at variance with studies of normal galaxies,
mock catalogues produced with these models are not publicly available.
For this reason, to test the reliability of equation (\ref{oldhon}), 
we are forced to follow an indirect approach by associating
quasar activity with a particular subset of synthetic galaxies.
In particular,
we consider galaxies which, at $z\simgt 1$, 
contain a substantial amount of cold gas in their nuclear region
that, in principle, could be accreted onto a central supermassive black hole.
Thus, in Appendix B, we use a publicly available
semianalytic model for galaxy formation (Hatton et al. 2003)
to study 
the halo occupation number of galaxies which, at $z\sim 1$, 
are actively forming stars in their bulges.
The reason for selecting this sample is threefold:
i) Even though imaged quasar hosts are consistent with
being massive ellipticals (e.g. Dunlop et al. 2003),
there is some observational evidence that, at high redshift, 
quasars might be associated with active star-formation (e.g. Omont et al. 
2001; Hutchings et al. 2002). 
ii) In the local Universe,
powerful Type 2 active galactic nuclei are found in bulges with either  
on-going star formation or young stellar populations
(Kauffmann et al. 2003).
iii) The observed correlation between black-hole mass and bulge 
velocity dispersion suggests that quasar activity and bulge
formation probably are physically associated phenomena. 
For instance, galaxy interactions and bar-induced inflows might funnel some
gas into the nuclear region of a galaxy thereby triggering 
simultaneous star formation and quasar activity. 

Our results show
that the halo occupation number of the simulated galaxies
is well approximated by a power-law with a rather sharp cutoff at low
masses.
This provides additional motivation to use
equation (\ref{oldhon}) in our analysis.

\subsection{Constraints from large-scale clustering}
\label{lss}

Assuming equation (\ref{oldhon}), we apply a least-squares method 
to determine the values of $M_0$ and $\alpha$ which best describe the 
clustering data presented in Fig. \ref{Fig:data}.
As discussed in Section \ref{results}, 
we use a principal component analysis of the errors
to deal with the clustering data.
We only consider  the region of parameter space
where $\alpha\geq 0$ and $M_0\geq 10^9\,M_\odot$.
We impose this lower limit to $\alpha$ since we expect the halo occupation
number to be a non-decreasing function of the halo mass.~\footnote{ 
Note that solutions with $\alpha< 0$ 
are allowed by the data if $M_0\sim 10^{12.5-13}
M_\odot$ (the precise value slightly depending on the assumed redshift range). 
In this case, quasars are hosted by haloes lying in a narrow mass range
which is centred around the values of $M_b$ given in Table \ref{Tab:PL}.
For this reason, there is no need to rediscuss these solutions here.}  
On the other hand, we assume a lower limit for $M_0$ because it is unreasonable
to consider halo masses which are smaller than the minimum inferred mass of 
the black holes powering our quasars (cf. Section \ref{bhm}).
In the range of separations covered by our dataset ($r\simgt 2$ \Mpc),
the two-halo term dominates 
the amplitude of the quasar 2-point correlation function   
and there is no need to specify the form of the function $\Sigma^2(M)$.
From equation (\ref{beff}), we also note that
the quasar correlation function on large scales
does not depend on the overall normalisation of $N(M)$ 
(e.g. the parameter $N_0$).

\begin{figure}
\centerline{\epsfig{figure=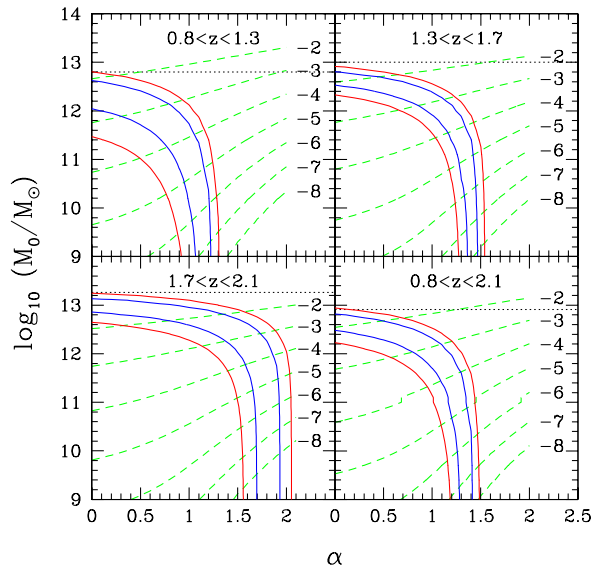,height=8cm}}
\caption{
Contour levels of the $\chi^2$ function for the parameters $\alpha$ and $M_0$ 
obtained by fitting
the large-scale clustering of quasars in different redshift intervals
with the predictions of the halo model given in equation (\ref{oldhon}).
The various panels contain
contour plots for the $\chi^2$ surface in the $\alpha-M_0$
plane. The contours correspond to $\Delta\chi^2=\chi^2-\chi^2_{\rm min}=1$ 
and 4 (respectively marking the 68.3 and 95.4 per cent
confidence levels for two fully degenerate Gaussian variables).
Contours of the value of $\log_{10} N_0$ which matches the observed 
number density are plotted as a function of $\alpha$ and $M_0$ (dashed
lines with labels) for the different redshift bins considered. 
The dotted lines mark the halo masses which correspond to 
the observed clustering
amplitudes (see Table \ref{Tab:PL}).
\label{Fig:chi_all}}
\end{figure}

Contour plots of the $\chi^2$ function are shown in Fig. \ref{Fig:chi_all}. 
Note that, independently of the redshift interval, the region of parameter
space allowed by the data is rather large and does not provide tight 
constraints on the values of $M_0$ and $\alpha$. This is because our data 
only fix 
the normalisation of the correlation function
(e.g. the bias parameter shown in Fig. \ref{Fig:bias}) 
and there is a whole one-dimensional family of $(\alpha,M_0)$ pairs which 
correspond to the same clustering amplitude. 

\subsection{Constraints from the number density} 
\label{dens}
In the halo-model described by equation (\ref{oldhon}), 
the number density of quasars 
depends on all the free parameters $\alpha$,
$M_0$ and $N_0$. In particular, $N_0$ acts as an overall normalising factor
so that, for given values of $\alpha$ and $M_0$, 
it can be expressed in terms of the mean number density of quasars
by combining equations (\ref{eq:avern}) and (\ref{oldhon}).
This gives an additional relationship among the parameters of the model
as, for every ($\alpha,M_0)$ pair, there is always a value of $N_0$ which 
exactly matches $n_{\rm QSO}$.
The dashed lines in Fig. \ref{Fig:chi_all} show the regions in the 
$(\alpha,M_0)$ plane where the observed density corresponds to a given 
value of $\log_{10} N_0$ (indicated by the labels in the plot).
The parameters $N_0$ and $M_0$ are strongly covariant: in order to get
the right quasar abundance, one needs to
lower the normalisation of the halo occupation number when $M_0$ is reduced.
The allowed range for $N_0$ spans many orders of magnitude,
reflecting the steep slope of the halo mass function at the low-mass end.

\subsection{Constraints from small-scale clustering}
\label{ssc}
For separations smaller than the typical size of the 
host haloes, the galaxy 
2-point correlation function is dominated by the contribution
of pairs lying within a single halo.
In this regime, $\xi(r)$ is fully described by equation (\ref{eq:xi1h})
which encodes information on the halo occupation distribution through its
second factorial moment, $\Sigma^2(M)$. 
This function can be conveniently expressed
in terms of the halo occupation number as follows:
\begin{equation}
\Sigma^2(M)=\Gamma(M)\, N(M)^2\;.
\label{esse}
\end{equation}
For a Poisson distribution, $\Gamma(M)=1$ independently of $M$.
In this case, measuring the 2-point correlation function 
on small scales provides additional constraints on the halo occupation number.

However, in general, the halo occupation distribution function is not 
Poissonian and $\Gamma(M)$ depends on the halo mass. 
In principle, this complicates
the estimate of $N(M)$ from analyses of small-scale clustering. 
In fact, a number of additional free-parameters might be
required to describe the behaviour of $\Gamma(M)$.
On the other hand, though, models of galaxy formation suggest that,
independently of the galaxy sample considered,
$\Gamma(M)$ is a very simple function which 
can be parameterized in terms of the same variables that are used
to describe $N(M)$.
Consistent results have been obtained for low-redshift galaxies
by using different semi-analytical models (e.g. Sheth \& Diaferio 2001; 
Berlind \& Weinberg 2002) and hydrodynamical simulations (Berlind et al. 2003).
Similarly, in Appendix B, we use a publicly available semianalytic model
to study the function $\Gamma(M)$ for star-forming galaxies at $z\simeq 1$.
In all cases, 
the scatter of $P_N(M)$ is strongly sub-Poissonian for haloes which, 
on average, contain less than 1 object, and nearly Poissonian for larger 
haloes.
This property plays a fundamental role in breaking the degeneracy among
all the models for the halo occupation number which can otherwise accurately 
describe galaxy clustering on large scales (Magliocchetti \& Porciani 2003).

We do not know whether the same conclusions apply to quasars. 
It is anyway interesting to understand what this would imply.
Let us assume that, also for quasars, $\Gamma\ll 1$ when $N\ll 1$
while $\Gamma\simeq 1$ when $N\simgt 1$.
Within the allowed parameter range in Fig. \ref{Fig:chi_all},
this implies  that, at variance with galaxies,
the 1-halo term never dominates the quasar correlation
function even on scales which are much smaller than the typical halo size.
This happens because the quasar $N(M)$ is always much smaller than unity and
its associated scatter is strongly sub-Poissonian. 
In other words,
the distribution of (optically bright) quasars in a halo is binary 
(either there is one or there is none) and
it is basically impossible to find 2 quasars being hosted by the same halo.

From the absence of quasar multiplets in a single halo it follows that, 
in order to use 
observational determinations of quasar clustering on
small scales to break the degeneracy among models which 
predict the same clustering amplitude on large scales,
one has to rely on the detection of halo exclusion effects (different haloes
cannot overlap). Note that this would be a direct ``measure'' of
the spatial dimension of dark-matter haloes and therefore of their mass.

The exact signature induced by spatial exclusion is hard to predict since dark 
matter haloes are expected to be triaxial objects and the precise form of the 
quasar correlation function
on small scales is also expected to depend on the 
position of each quasar within its host halo 
(see equation \ref{eq:xi1h}). However, it is clear that the configuration
which maximizes this effect is obtained when quasars sit at the centre
of the haloes. In this case, the distribution of quasars is a perfect
(sparse sampled)
tracer of the underlying halo distribution and the 2-point correlation
function, $\xi(r)$, is expected to reach the value -1 on scales smaller
than the typical size of the host halo.
Such exclusion effects will then correspond to a flattening
of the projected function $\Xi$ on the same scales.
However, these effects might be hard to detect due to the 
small number statistics of close quasar pairs (cf. Fig. \ref{Fig:data}).

Similar arguments apply to any population of rare objects. 
A possible detection (with relatively low statistical significance) 
of exclusion effects has been reported from 
the analysis of the clustering properties of 
Lyman-break galaxies at redshift $\sim$3 (Porciani \& Giavalisco 2002).

Note, however, that the scatter of the halo occupation distribution might
depend
on the detailed physical processes which give rise to the quasar phenomenon
and thus be very different from the $\Gamma$ function which describes the 
galaxy distribution. Therefore, the presence of quasar multiplets
inside single haloes is not ruled out by the data.
Measuring the quasar 2-point correlation function on separations
smaller than 1 \Mpc\ would give a definitive answer to this question.

\section{The halo occupation number from quasar luminosities}
\label{sol}
Recent studies of stellar and gas dynamics in local galaxies have revealed
a wealth of information on the population of supermassive black holes.
The observational evidence for a correlation between the
mass of a black hole, $M_{\rm bh}$, 
and the circular velocity, $v_c$, of its host galaxy
(Ferrarese 2002; Baes et al. 2003) is one of the most intriguing results. 
In this section, 
we use this empirically determined relation
($M_{\rm bh}\propto v_c^{4.2}$) to derive the halo occupation number
of quasars in the 2QZ.
This is obtained by first converting quasar luminosities into a distribution
of black-hole masses (to which we apply the $M_{\rm bh}-v_c$ correlation)
and then linking, with minimal assumptions, 
the circular velocity of the host 
galaxies to the mass of their dark-matter haloes.

\subsection{The mass of quasar host haloes}
\label{revmass}
The bolometric luminosity of a quasar and the mass of the accreting black
hole can be related as follows
\begin{equation}
\frac{M_{\rm bh}}{M_\odot}=
\frac{1}{\eta}\,
\frac{L_{\rm bol}}{1.26\times 10^{38}\,{\rm erg}\,{\rm s}^{-1}}
\label{Mbh}
\end{equation}
where $\eta$ denotes the ratio between the bolometric luminosity of the quasar
and the Eddington luminosity.
We use this relation to determine the function ${\cal P}(M| M_{b_{\rm J}})$
which gives 
the conditional probability distribution of the host-halo mass, $M$, 
for a quasar with given absolute magnitude $M_{b_{\rm J}}$.
For ease of reading, we just summarize our calculations here while
a detailed presentation of the model is given in Appendix A.

To compute ${\cal P}(M| M_{b_{\rm J}})$,
 we first determine the conditional probability of 
$M_{\rm bh}$ given $M_{b_{\rm J}}$.
This is obtained by combining the most recent bolometric corrections
from the B band with an empirically determined 
distribution of Eddington ratios (McLure \& Dunlop 2004). 
We then use the observed $M_{\rm bh}-v_{\rm c}$ relation 
(Baes et al. 2003) to derive the conditional probability of $v_{\rm c}$
given $M_{b_{\rm J}}$.
Eventually, we convert circular velocities into halo masses by 
assuming that the observed circular velocity and the virial velocity 
of the host halo are related by $v_{\rm c}=\psi\, v_{\rm vir}$
with $\psi$ a free parameter.
Recent lensing studies (Seljak 2002) suggest that, in the mass range
of interest, $\psi=1.4\pm 0.2$ (case A); alternatively, theoretical arguments
based on the estimated low concentration of high-redshift haloes, 
suggest that $\psi\simeq 1$ (case B).
These different choices bracket the range of plausible values for $\psi$
(see the detailed discussion in Appendix A).

\subsection{The halo occupation number}
\label{lumhon}

\begin{figure}
\centerline{
\epsfig{figure=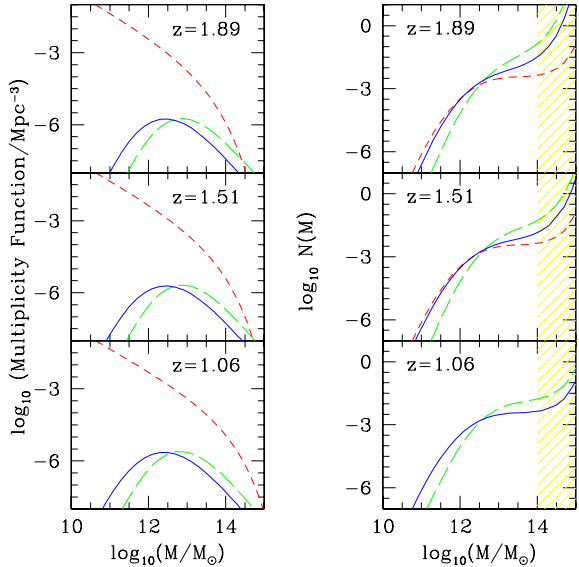,height=8cm}}
\caption{Left panel: the multiplicity function 
(differential number density per log-unit of mass and per unit volume)
of haloes which are hosts of 2QZ quasars (continuous
line for case A, long-dashed line for case B) is compared with the 
corresponding distribution for all the dark-matter haloes (short-dashed line) 
at different redshifts. 
Right panel: corresponding halo occupation numbers obtained by
taking the ratio of the quasar-host mass function to the total halo 
mass function (continuous line for case A, long-dashed line for case B).
For comparison, the case A solution at $z=1.06$ is reproduced with
a short-dashed line in the two top panels.
\label{Fig:lf-hon}}
\end{figure}

In this section, we test
whether the conditional mass distribution ${\cal P}(M| M_{b_{\rm J}})$
is consistent with the quasar 
clustering data we measured in Section \ref{results}.
By integrating over the luminosity function,
${\cal P}(M| M_{b_{\rm J}})$
can be easily turned into  
the mass function of dark haloes which are quasar hosts,
\begin{equation}
n_{\rm q}(M,z)
=\int_{M^{\rm f}_{b_{\rm J}}(z)}^{M^{\rm b}_{b_{\rm J}}(z)}
\Phi(M_{b_{\rm J}},z)\,{\cal P}(M| M_{b_{\rm J}})\,{\rm d}M_{b_{\rm J}}\;.
\end{equation}
The corresponding multiplicity function, $M\, n_q(M,z)$,
is shown in the left panel of Fig. \ref{Fig:lf-hon} for different
values of $z$. 
Independently of redshift, the halo mass distribution per unit
logarithmic interval of $M$ peaks 
around $10^{12.5-13} M_\odot$:
this is the characteristic mass of quasar hosts, whose value is comparable 
with those listed in Table \ref{Tab:PL}.
By then dividing $n_{\rm q}(M,z)$ by the halo mass function,
we obtain a new estimate for the halo occupation number
\begin{equation}
N(M,z)=\frac{n_{\rm q}(M,z)}{n(M,z)}\;.
\label{newn}
\end{equation}
Note that this, in principle,
could give rise to a biased estimate of $N(M,z)$.
In fact,
in a CDM cosmology, each virialized halo contains a number of sub-haloes
within its $r_{\rm vir}$, and at least some of these sub-haloes will be 
associated with
galaxies which formed within their local overdensities and then falled
into the larger halo. Since the rotational
properties of galaxies are expected to be related to local dark-matter 
overdensities,
the values we derived for $M$ most likely refer to 
sub-haloes. On the other hand, 
$n(M)$ describes the mass distribution of the parent haloes.
We can account for this problem by introducing the conditional 
mass function 
of the sub-haloes of mass $M_{\rm s}$ which lie  within a parent halo of mass 
$M_{\rm p}>M_{\rm s}$, $n(M_{\rm s}|M_{\rm p})$. 
The probability that an halo with mass $M$ is a parent one is then given by
\begin{equation}
{\cal P}_{\rm p}(M)=
\frac{n(M)}{n(M)+
\int_{M}^\infty n(M_{\rm p})\,n(M|M_{\rm p})\,{\rm d}M_{\rm p}}\;,
\end{equation}
where the integral at the denominator
gives the mass function for the sub-haloes.
We use two different functional forms for $n(M_{\rm s}|M_{\rm p})$
which have been derived from high-resolution numerical simulations
(Sheth \& Jain 2003; Vale \& Ostriker 2004 and references
therein).
In both cases, we find that the sub-halo correction is negligibly small.
In fact, at the redshifts spun by our quasar sample, 
the haloes we are considering are rather massive 
and only a few per cent of them have been included into larger units.

Results for the halo occupation number are  
presented in Fig. \ref{Fig:lf-hon}. 
In all cases, for $M<10^{14} M_\odot$,
$N(M)$ is well approximated by a broken power law
which qualitatively resembles equation (\ref{oldhon}). The low-mass tail
has a typical slope $\sim -3.6$ for case A and $\sim -4.2$ for case B, 
independently of redshift.
On the other hand, the high-mass tail gets progressively steeper
when moving from low to high redshifts. For case A, we get a flat $N(M)$
for the low-redshift sample, to be compared with a slope of $\sim 0.4$ 
for the dataset at intermediate redshifts
and of $\sim 0.7$ for the highest-redshift quasars.
The corresponding numbers for case B are $\sim 0.4$, $\sim 0.9$, $\sim 1.1$.
For $M>10^{14} M_\odot$, in both cases the halo occupation number
starts growing exponentially. This happens because, in the high-mass tail,
$n_{\rm q}(M,z)$ does not drop as fast as 
the exponential cutoff of the halo mass function
(see the left panel in Fig. \ref{Fig:lf-hon} and the shaded region in
the right panel). 
Most likely this is a spurious effect due to the simple assumptions
we use to derive $n_{\rm q}(M,z)$. This artifact, however, does not affect our 
conclusions since the fraction of quasars that are found to reside in
such massive haloes is extremely small
(at most, 
a few $\times 10^{-5}$ for case A and less than 2 per cent for case B).

Note that estimates for $N(M)$ based on equation (\ref{newn}) are obtained
without any information on the clustering properties of quasars.
It is therefore interesting to check 
whether they
are in agreement with the determination of $\Xi^{\rm obs}(r_\perp)$ 
we presented in Section \ref{results}. 
For this reason, 
we compute the effective bias associated with the different
halo occupation numbers presented in Fig. \ref{Fig:lf-hon}. 
For the lowest-redshift sample, 
we get $b_{\rm eff}=1.63$ (2.02) for case A (B).
At intermediate redshifts,
we find $b_{\rm eff}=2.14$ (2.73) for case A (B).
while, for the highest-redshift interval,
we derive $b_{\rm eff}=2.58$ (3.36) for case A (B).
These numbers have to 
be compared with the observational results presented in Table \ref{Tab:PL}. 
For case A, we note that, even though the estimated bias parameter
increases with redshift (as like as the data in Table \ref{Tab:PL}),
its value is in general too low to accurately describe the 
observed clustering. In other words, case A
tends to underestimate the mean mass of quasar host haloes.
Predictions for case B, instead, are of better quality.
In this case, our results for the bias parameter are  
rather accurate for the intermediate redshift sample,
while they tend to overestimate (underestimate) the
observational results at low (high) redshifts.
Anyway, the bias inferred from our models is always 
acceptable (with respect to the 
statistical errors associated with the determinations of 
the projected correlation function).
The maximum discrepancy appears at $z_{\rm eff}=1.89$ and
corresponds to a statistical significance of $1.4\, \sigma$.
These results are in agreement with the recent analysis by 
Wyithe \& Loeb (2004)
who showed that quasar models with $M_{\rm bh}\propto v_{\rm c}^5$,
$\psi=1$ and $\eta\sim 0.1-1$ are able to reproduce the evolution of the
correlation length measured in a preliminary data release 
of the 2QZ (Croom et al. 2001, 2002).
More accurate clustering measurements are then required to detect possible
changes in the $M_{\rm bh}-v_c$ correlation and to distinguish them
from effects due to evolution of other parameters
(for instance $\eta$, $\psi$ or the bolometric correction).

\begin{figure*}
\centerline{\epsfig{figure=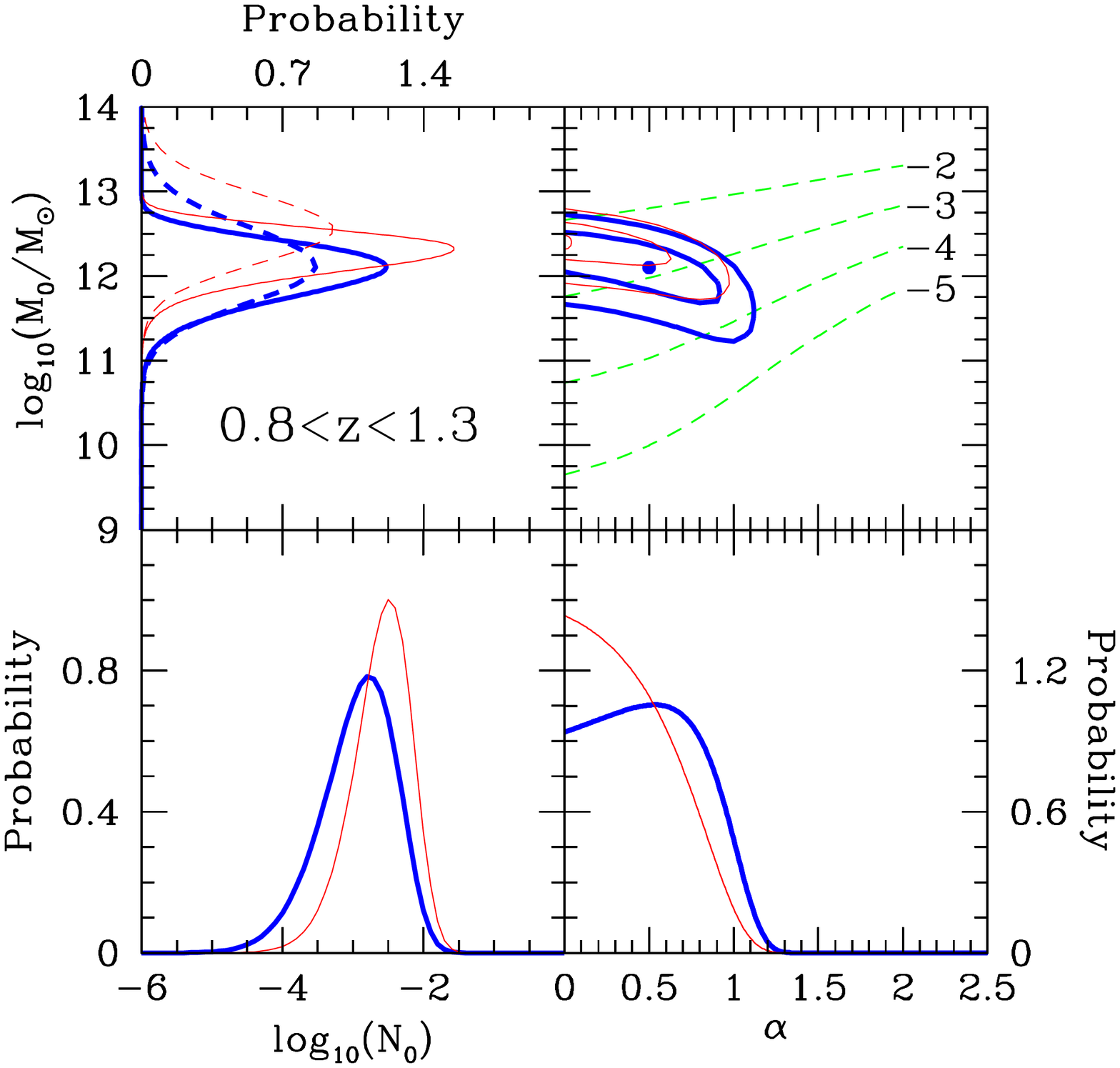,height=7.5cm}
\epsfig{figure=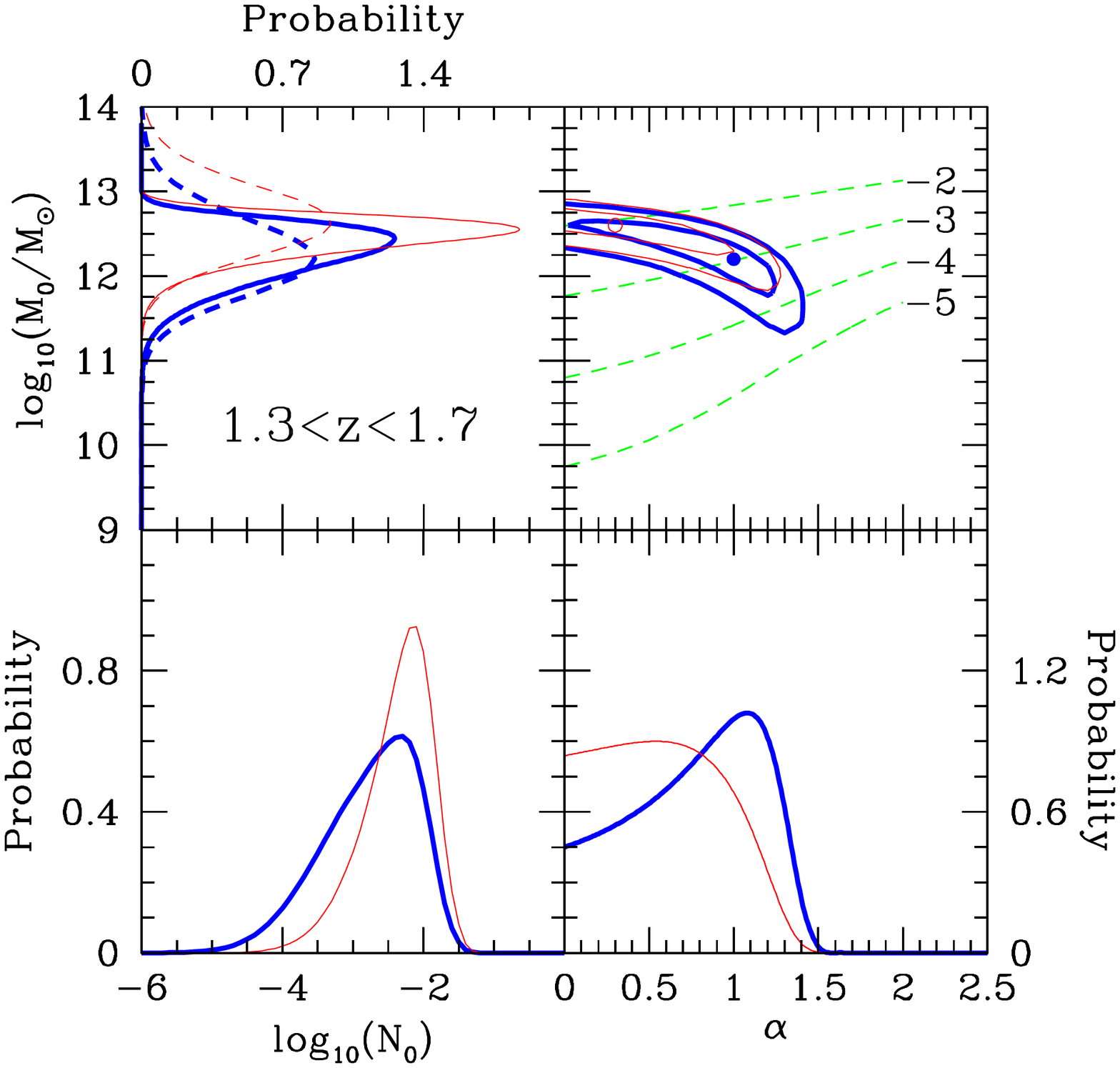,height=7.5cm}}
\centerline{\epsfig{figure=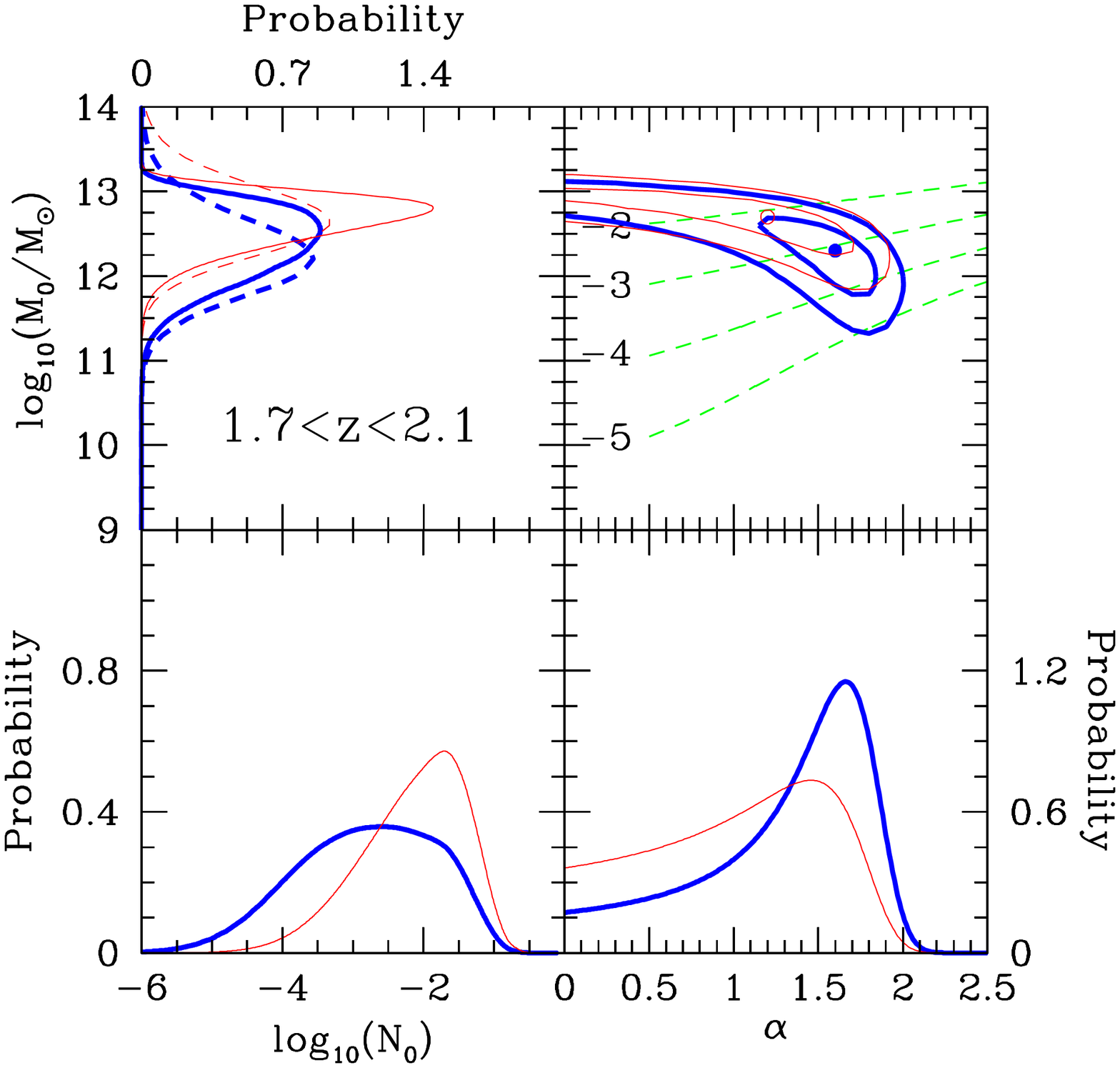,height=7.5cm}
\epsfig{figure=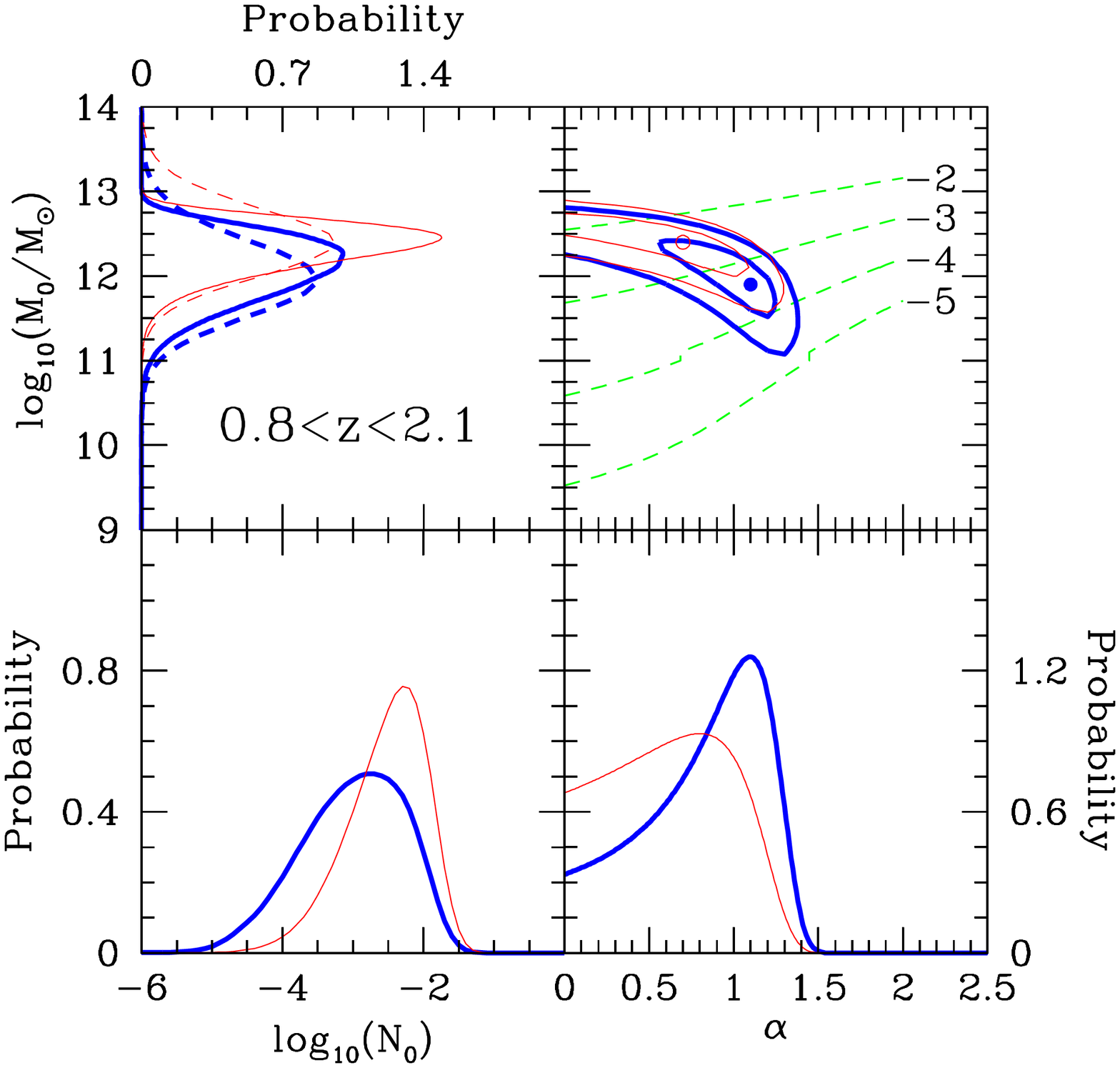,height=7.5cm}}
\caption{Posterior probability distribution for the parameters of the halo
model. The four frames correspond to different 
redshift ranges as indicated by the labels. 
Thick and thin lines respectively refer to Prior A and B.
{\it Top-right:}
Contours of the joint distribution of $\alpha$ and $M_0$ 
(obtained by marginalizing the three-dimensional posterior probability 
over $N_0$). The most-probable point is marked with a small circle.
To facilitate the comparison with Fig. \ref{Fig:chi_all}, 
the continuous lines 
show the points where $-0.5\ln {P_{\rm max}/P}=1$ and 4
(which, in this case, do not have any special meaning).
As in Fig. \ref{Fig:chi_all}, in order to represent the covariance of the
different parameters,
the dashed lines show the loci in the ($\alpha,M_0$) plane where
a given value of $\log_{10} N_0$ (indicated by the labels)
perfectly matches the observed number density of quasars.
{\it Other panels:}
The probability densities for each single parameter
(obtained by marginalizing the posterior distribution
over the remaining two variables) are shown in the top-left (for $\log_{10} M_0/M_\odot$),
bottom-left (for $\log_{10} N_0$) 
and bottom-right (for $\alpha$) panels. In the top-left panels, 
the dashed lines show the assumed prior distributions for $M_0$.
\label{prior}}
\end{figure*} 
\begin{table*}
\begin{center}
\caption{Best-fitting values for the parameters of the halo model
for different redshift ranges (superscript bf).
The last three columns list the central 68.3 per cent credibility intervals 
for each single parameter. 
These have been 
obtained by marginalizing the posterior probability distribution function 
over the remaining parameters. The quoted values correspond to 
the 15.85, 50 and 84.15 percentiles.
In the last column, we list the 
central 90 per cent credibility intervals for the quasar lifetime.
We assume that a halo of $2\times 10^{13} M_\odot$
harbours, on average, a single super-massive black hole so that
the halo occupation number of bright quasars coincides with their
duty cycle.
\label{Tab:CLprior}} 
\begin{tabular}{cccccccccc}
\hline
\hline
$z_{\rm min}$&$z_{\rm max}$ & Prior & $\log_{10}(M_0^{\rm bf}/M_\odot)$ &
$\alpha^{\rm bf} $ & 
$\log_{10} N_0^{\rm bf}$ &
$\log_{10}(M_0/M_\odot)$ &
$\alpha $ & 
$\log_{10} N_0$ & $t_{\rm Q}/10^7$ yr\\
\hline
$0.8$&$1.3$ & A & 12.1 & 0.5 & -2.9 &
$12.1^{+0.3}_{-0.4}$ & $0.5^{+0.3}_{-0.3}$ & $-2.9\,^{+0.4}_{-0.6}$&
$1.9^{+1.8}_{-1.3}$\vspace*{0.1cm}\\ 
$0.8$&$1.3$ & B & 12.4 & 0.0 & -2.3 & 
$12.2^{+0.2}_{-0.3}$ & $0.4^{+0.3}_{-0.3}$ & $-2.6^{+0.3}_{-0.5}$&
$2.3^{+1.9}_{-1.4}$\vspace*{0.1cm}\\ 
\hline
$1.3$&$1.7$ & A & 12.2 & 1.0 & -3.0 & 
$12.3^{+0.3}_{-0.4}$ & $0.8^{+0.4}_{-0.5}$ & $-2.7^{+0.6}_{-0.8}$&
$3.2^{+1.7}_{-1.6}$\vspace*{0.1cm}\\
$1.3$&$1.7$ & B & 12.6 & 0.3 & -2.1 & 
$12.5^{+0.2}_{-0.3}$ & $0.6^{+0.4}_{-0.4}$ & $-2.3\,^{+0.3}_{-0.6}$&
$3.2^{+2.3}_{-1.5}$\vspace*{0.1cm}\\
\hline
$1.7$&$2.1$ & A & 12.3 & 1.6 & -3.1 & 
$12.4^{+0.4}_{-0.5}$ & $1.4^{+0.4}_{-0.7}$ & $-2.8^{+1.0}_{-1.1}$&
$5.9^{+5.3}_{-2.4}$\vspace*{0.1cm}\\
$1.7$&$2.1$ & B & 12.7 & 1.2 & -2.1 & 
$12.7^{+0.3}_{-0.4}$ & $1.1^{+0.5}_{-0.7}$ & $-2.1^{+0.7}_{-0.8}$&
$6.8^{+7.3}_{-2.9}$\vspace*{0.1cm}\\
\hline
$0.8$&$2.1$ & A & 12.0 & 1.0 & -3.1 & 
$12.1^{+0.3}_{-0.4}$ & $0.8^{+0.3}_{-0.5}$ & $-2.8^{+0.6}_{-0.8}$&
$3.0^{+2.2}_{-1.8}$\vspace*{0.1cm}\\
$0.8$&$2.1$ & B & 12.4 & 0.5 & -2.2 & 
$12.3^{+0.2}_{-0.3}$ & $0.6^{+0.4}_{-0.4}$ & $-2.4^{+0.4}_{-0.6}$&
$3.2^{+2.6}_{-1.9}$\vspace*{0.1cm}\\
\hline
\end{tabular}
\end{center}
\end{table*}

\section{A Bayesian analysis of the halo occupation number}
\label{bayes}
We have shown that, at variance with galaxy clustering, lack of information
on (and from) the 2-point correlation function of quasars on small scales 
does not allow us to break the degeneracy among the best-fitting models
presented in Fig. \ref{Fig:chi_all}.
In this section, we adopt a Bayesian approach and use information on 
quasar luminosities to further constrain the parameters of the halo
occupation number.

Assuming equation (\ref{oldhon}), for each redshift interval,
we translate the probability density function for
the errors of $n_{\rm QSO}$ and $\Xi(r_\perp)$
into a likelihood function for the model parameters 
and we write
\begin{equation}
{\cal L}_{\rm tot}(\alpha,M_0,N_0)={\cal L}_{\rm clust}(\alpha,M_0)\cdot
{\cal L}_{\rm dens}(\alpha,M_0,N_0)\;.
\label{ltot}
\end{equation}
where ${\cal L}_{\rm clust}$ accounts for the large-scale clustering analysis
presented in Section \ref{lss} and ${\cal L}_{\rm dens}$ for the 
quasar number density.~\footnote{Assuming Gaussian errors, 
$-2 \ln {\cal L}_i=\chi^2_i + {\rm const}$ where $\chi^2_i$ denotes
the usual chi-square statistic.}
In other words, the number density constraints weights as much as a single
independent point in the clustering analysis.
We then apply Bayes' theorem to our dataset.
For simplicity, to express our lack of prior knowledge,
we adopt constant (non-informative) prior distributions
for $\log_{10} N_0$ and $\alpha$.
On the other hand, as a prior distribution for $\log_{10} (M_0/M_\odot)$,
we use the probability distribution ${\cal P}(M_0| M_{b_{\rm J}}^{\rm max})$ 
that we derived in Appendix A and presented in Section \ref{revmass}.
This is the probability
distribution of the halo masses which harbour the faintest quasars
that can be detected in each 2QZ sample considered.
This prior knowledge is based on the empirically
determined correlation between black-hole masses and the circular velocity
of the host galaxies.
%
%
As previously discussed,
to test the robustness of our method with respect to underlying
systematic uncertainties,
we consider two different prior distributions corresponding to  
$\psi=1.4\pm0.2$ (Prior A) and $\psi=1$ (Prior B).

Contours of the posterior probability in the $\alpha-M_0$ plane
and the probability distribution of the single parameters (marginalized
over the remaining ones) are shown in Fig. \ref{prior}.
The corresponding best-fitting values and credibility intervals for the
different parameters are listed in Table \ref{Tab:CLprior}.

Note that adopting 
our informative prior on $M_0$ is enough to break the degeneracy among
the parameters of the best-fitting models.
In practice, both priors exclude the region  $M_0<10^{11} M_\odot$ where
haloes are too small to harbour bright quasars.
This is sufficient to determine a non-degenerate solution for each
redshift range.
The main characteristics of these solutions can be summarized as follows.
In general, 
the cutoff mass, $M_0$, has a very mild evolution with redshift. 
Using prior A, we get $M_0\sim (1-3) \times 10^{12} M_\odot$,
while, with prior B, we obtain $M_0\sim (2-5) \times 10^{12} M_\odot$.~\footnote{Note that $M_0$ is only mildly covariant with the high-mass slope, 
$\alpha$,
in the sense that slightly lower values for $M_0$ are generally associated with
larger values of $\alpha$.}
On the other hand, in order to match the rapidly evolving bias parameter
of the three quasar samples with a nearly invariant $M_0$,
the high-mass slope, $\alpha$, tend to become steeper and 
steeper with increasing $z$. 
This is in qualitative agreement with the results presented in Section
\ref{lumhon}.
At $z_{\rm eff}=1.06,\ 1.51$ and 1.89, we respectively find
$\alpha\sim 0.5,\ 0.8$ and $1.4$ for prior A, 
and $\alpha\sim 0.4,\ 0.6$ and $1.1$ for prior B.
It is important to stress, however, that
the parameter $\alpha$ is typically poorly determined.
Strictly speaking, the data just set an upper limit for it.
The allowed range for the normalisation parameter $N_0$
varies systematically with the assumed prior. 
In brief, $N_0$ spans a broader range (approximately 
from $3\times 10^{-5}$ to $2\times 10^{-2}$) when prior A is used.
On the other hand, with prior B, the probability distribution
for $N_0$ is more peaked and ranges from $3\times 10^{-4}$ to $
3\times 10^{-2}$.
We note that, for both priors, 
$N_0$ is less tightly determined for our high-redshift
sub-sample.

\section{The quasar lifetime}
\label{lifetime}
The number of optically bright quasars per halo can be used to estimate
the duty cycle of quasar activity
and, thus, the quasar lifetime (Haiman \& Hui 2001; Martini \& Weinberg 2001).
In brief, let us assume, for simplicity, that each dark-matter halo contains
one supermassive black-hole. In this case, the fraction of active quasars 
per halo coincides with the quasar duty cycle. Assuming that quasar activity
is randomly triggered (for instance by tidal interactions with neighbours)
during the halo lifetime, $t_{\rm H}$,
the duty cycle can then be expressed as $t_{\rm Q}/t_{\rm H}$
where $t_{\rm Q}$ denotes a characteristic timescale over which the quasar
is visible in the optical band. Both 
a single optically bright phase and a series of shorter bursts correspond
to the same $t_{\rm Q}$. 
\subsection{Estimating the quasar lifetime}
\begin{figure}
\epsfig{figure=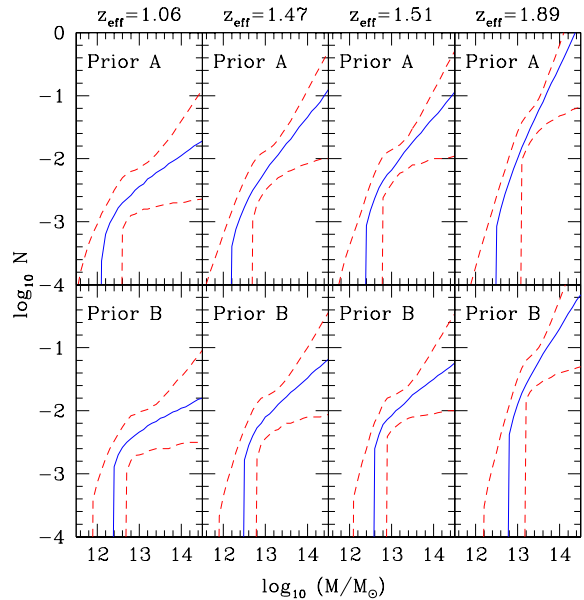,height=8cm}
\caption{Halo occupation number obtained from the posterior
probability distribution in Fig. \ref{prior}.
For each halo mass, $M$, 
we derive the probability density function of $N$ using equation (\ref{npdf}). 
Solid lines show the median occupation number as a function of $M$ while
dashed lines indicate the 5 and 95 percentiles of the distribution. 
\label{Fig:tq}}
\end{figure}

In this section, we use 
the posterior probability distribution presented in  Fig. \ref{prior}
to determine the characteristic quasar lifetime.
For each halo mass,
we first compute the probability density function of the halo occupation 
number:
\begin{equation}
{\cal P}(N)=\int \delta[N-N(M;\theta)]\,P(\theta|{\bf D})\,
{\rm d}^3\theta
\label{npdf}
\end{equation}
where $\delta(x)$ denotes the Dirac-delta distribution,
$\theta\equiv(\alpha,M_0,N_0)$,
$P(\theta|{\bf D})$ is the posterior probability  and 
$N(M;\theta)$ is given in equation (\ref{oldhon}).
The 5, 50 and 95 percentiles of this distribution are shown in
Fig. \ref{Fig:tq} as a function of the halo mass. 
Note that results are nearly independent of the considered prior.
For haloes with $M\simeq 10^{13} M_\odot$, the occupation number is
tightly constrained by the data.
We thus estimate the characteristic quasar lifetime by assuming
that, for these haloes, $N=t_{\rm Q}/t_{\rm H}$.
Following Martini \& Weinberg (2001),
the halo lifetime is defined as the median time interval during which 
a halo of mass $M$ at redshift $z$ is incorporated into a halo of mass $2M$.
This quantity is computed using equation (2.22) of Lacey \& Cole (1993).

Results for $t_{\rm Q}$ are listed in Table  \ref{Tab:CLprior}.
We find that the estimated quasar duty cycle increases with $z$ 
(and/or with quasar luminosity).
%
For our sample at $z_{\rm eff}=1.06$, we find that
only $\sim 0.6$ per cent of the host-haloes
with $M=2\times 10^{12} M_\odot$
contain a bright quasar, which corresponds
to $t_{\rm Q}\simeq 2\times 10^{7}$ yr.
This coincides with the $e$-folding time of a black hole
which accretes mass with a radiative efficiency $\epsilon \sim 0.1$
and shines at a fraction $\eta\sim 0.5$ of its Eddington luminosity
(Salpeter 1964).
On the other hand, at $z_{\rm eff}=1.89$,
the fraction of active black-holes increases to $\sim 5$ per cent
and $t_{\rm Q}\simeq 7\times 10^7$ yr.
In all cases, the estimated lifetime lies between $10^7$ and $10^8$ yr.

Even though the determination of $N(M)$ becomes more uncertain 
for $M\gg 10^{13} M_\odot$, our results suggest that the occupation
number of quasars tends to increase with the halo mass.
This, however, does not imply that $t_{\rm Q}$ augments as well.
In fact, 
our estimates for the quasar lifetime are degenerate with the occupation number
of supermassive black holes which, most likely, increases with the halo mass.

Note that, given the simplicity of the model, our
results are only indicative. The quoted quasar lifetimes should be revised
upwards if: (i) a non-negligible number of haloes do not harbour any 
supermassive black-hole; (ii) optical radiation from quasars turns out to be 
significantly beamed; (iii) in the presence of an important fraction of 
obscured sources.
On the other hand, $t_{\rm Q}$ is smaller than what reported here 
if more than one supermassive black-hole is hosted, on average, by each halo.

A number of observations hint towards a one to one correspondence between
supermassive black holes and host haloes.
High-resolution optical imaging with the {\it Hubble Space Telescope} (HST)
shows that bright quasars ($M_V< -23$) at $z<0.5$
are only harboured by
exceptionally luminous galaxies with $L \simgt L^*_V$
(Bahcall et al. 1997; Hamilton, Casertano \& Turnshek 2002).
These galaxies turn out to be a mixture of different morphological
types, ranging from normal ellipticals and spirals to complex systems
of gravitationally interacting components (Bahcall et al. 1997).
However, a number of observational results suggest that
spheroidal hosts become more prevalent with increasing nuclear
luminosities: quasars with $M_V < -23.5$ are virtually all
harboured by luminous elliptical galaxies (Dunlop et al. 2003).
Similarly, observed surface-brightness profiles
suggest that bright quasars at $z\sim 1-2$
are hosted by massive ellipticals undergoing passive
evolution (Kukula et al. 2001; Falomo et al. 2004).~\footnote{
Some authors, however, find that a disc-like component is always needed
to accurately fit the data at large radii
(Percival et al. 2001; Hutchings et al. 2002).} 
%
%
%
Taking this for granted, we can show that our assumption of one supermassive
black hole per halo (and thus our inferred quasar lifetime) is rather
realistic. The argument proceeds as follows.
(i)
Massive elliptical galaxies are made of old stellar populations which formed 
at $z\simgt 2$ and passively evolved thereafter. 
(ii) In the assumed cosmology, the massive haloes which harbour these galaxies 
can only increase 
their mass by a factor of a few from $z=1-2$ to the present epoch.
This mainly happens via accretion of smaller objects.
(iii) From clustering studies in the local Universe, we derive that
haloes with $M\sim 10^{13-14} M_\odot$ harbour on average
$\sim 1-2$ early-type galaxies
with effective luminosity $L_{\rm eff}=1.3\, L^*$ 
(Magliocchetti \& Porciani 2003).
Points (i), (ii) and (iii) imply
that the mean number of early-type galaxies per halo 
was of order unity even at $z\sim 1-2$.
Thus, our thesis follows from the assumption that
each galaxy hosts a single supermassive black hole.

%
%

%

\subsection{Constraints from the proximity effect}
A quasar produces 
enhanced ionisation of H and He in its surroundings 
thus creating opacity gaps in its spectrum (or in the spectrum of   
background QSOs lying on adjacent lines of sight).
The physical characteristics of these HII and HeIII regions can be used to 
estimate the
quasar lifetime (e.g. Bajtlik, Duncan \& Ostriker 1988; Heap et al. 2000).
A number of studies have shown that,
in order to explain the proximity effect in the Ly-$\alpha$ forest,
quasars have to maintain their ionizing luminosity for at least $10^5$ yr
(e.g. Bajtlik, Duncan \& Ostriker 1988; Schirber et al. 2004).
At the same time, the best estimates for the quasar lifetimes indicate that
$t_{\rm Q}\simgt 10^7$ yr  (Hogan et al. 1997; Anderson et al. 1999; 
Jakobsen et al. 2003).
It is interesting to check what are the implications of these observational
results for our halo model.
We first note that extremely short quasar lifetimes correspond
to very low values for $N_0$. From Fig. \ref{Fig:chi_all}, we then learn
that the condition $t_{\rm Q}> 10^5$ yr
basically rules out all the models with $M_0\simlt 10^{11} M_\odot$.
On the other hand, a timescale $\simgt 10^7$ yr is 
is fully consistent with our results for $t_{\rm Q}$ presented
in Section \ref{lifetime}.
In other words, constraints to the halo model from quasar luminosities
and from the proximity effect are consistent and approximately equivalent. 
Future determinations of quasar radiative histories 
based on the transverse proximity effect (e.g. Adelberger 2004 and references
therein) will hopefully provide more stringent limits.

\section{Is the halo occupation number evolving?}
\label{evo}
The analysis presented in Section \ref{sechon} is based on two basic assumptions:
the halo model and an assumed functional form for $N(M)$,
namely, equation (\ref{oldhon}). 
Within these working hypotheses,
in the three-dimensional parameter space $(\alpha,M_0,N_0)$ we identified
a one-dimensional family of models which 
accurately fits the abundance and clustering properties of quasars in 
the 2QZ. Prior information on $M_0$,  
inferred from quasar luminosities, was used in Section \ref{sol}
to remove the degeneracy between
the model parameters. 

In this section, we want to test whether quasar clustering
(without any additional constraint from quasar luminosity)
is consistent with a non-evolving model for $N(M)$. 
Indeed, the contours in Fig. \ref{Fig:chi_all} obtained  
for quasars at different redshfits tend to lie in the same region of 
the $\alpha-M_0$ parameter space. 
This also applies to the
halo occupation number of the entire quasar sample ($0.8<z<2.1$).
We then assume, as a working hypothesis, that the shape of the halo occupation 
number (parameterized by $\alpha$ and $M_0$)
does not evolve within the time-interval spun by our quasar dataset. 
On the other
hand, we let the overall normalisation $N_0$ vary. In fact, because of
selection effects, quasars lying at higher redshifts tend to be (on average) 
intrinsically brighter than their lower-redshift counterparts 
(cf. Table \ref{Tab:data}).
Therefore, since we are considering objects within different luminosity
ranges, it is reasonable to assume that they will correspond to 
different values of $N_0$ and, thus, to different number densities.
We denote these new parameters as $N_0^{\rm L}$,
$N_0^{\rm M}$ and $N_0^{\rm H}$ respectively for the low, median
and high redshift samples. 

\begin{figure}
\epsfig{figure=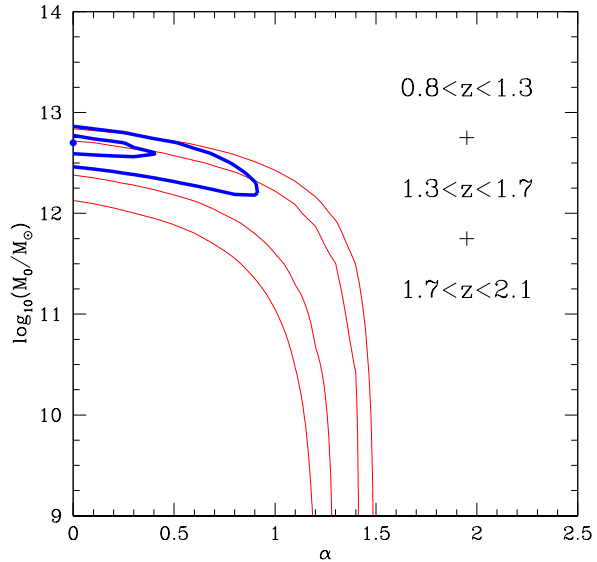,height=8cm}
\caption{
Contour levels for the $\chi^2$ function obtained by combining
the quasar sub-samples at different redshifts and
assuming that the shape of the halo occupation number does not evolve
with $z$.
The $\chi^2$ function is shown as a function of the   
halo model parameters $\alpha$ and $M_0$ and it has been minimized
with respect to $N_0^{\rm L}, N_0^{\rm M}$ and $N_0^{\rm H}$.
A point indicates
the best-fitting model while the heavy lines mark the 
68.3 and 95.4 per cent confidence levels
(respectively corresponding to $\Delta\chi^2=2.3$ and 6.17).
For ease of comparison, the contours 
presented in the bottom right panel of Fig. \ref{Fig:chi_all} 
are represented with light lines. These refer to the halo occupation number
of our entire quasar sample in the redshift range $0.8<z<2.1$.
\label{Fig:combo}}
\end{figure}

In Fig. \ref{Fig:combo}, we show the confidence levels in the 
$(\alpha,M_0)$ plane obtained by combining 
the three redshift subsamples.
The objective function (total $\chi^2$) has been computed by adding together
the $\chi^2$s of each sample. The contours shown in the figure are obtained
by minimizing the total $\chi^2$ function over the different $N_0^i$s 
(we remind the reader that for each pair of values for $(\alpha,M_0)$ it
is always possible to choose the $N_0^i$s 
so that to perfectly match the observed densities).

The minimum value assumed by the total $\chi^2$ function
over the parameter space is 10.98 with 10 degrees of freedom.
Therefore, assuming Gaussian errors,
our working hypothesis that the halo occupation number of bright quasars
does not evolve with redshift is not rejected by the data at any significant
confidence level.
The best fitting values for the parameters are: 
$\alpha=0.0^{+0.4}$, $M_0=12.7\pm 0.1$, 
$\log_{10}N_0^{\rm L}=-1.96^{+0.14}_{-0.26}$,
$\log_{10}N_0^{\rm M}=-1.86^{+0.15}_{-0.24}$
and $\log_{10}N_0^{\rm H}=-1.73^{+0.17}_{-0.27}$.
All the quoted intervals correspond to $\Delta \chi^2=1$.
This corresponds to $b_{\rm eff}=2.08\pm 0.10$, $2.64\pm 0.15$ and 
$3.20\pm 0.20$
respectively for the low, medium and high redshift samples.
In this case then, changes in the bias parameter are merely driven by the
joint time evolution of the halo population and of the mass autocorrelation 
function. 

In summary, the combined dataset is consistent with a model
for the halo occupation number which does not evolve with lookback time 
and exhibits a very shallow dependence on the halo mass
($\alpha<1$, with values near zero which are favoured). One also finds
$M_0\simeq 5\times 10^{12} M_\odot$ and $N_0^i\simeq 0.01-0.02$.
For all the quasar sub-samples,
this corresponds to $t_{\rm Q}\simeq (3-4) \times 10^7$ yr.

\section{Discussion}
\label{discuss}
\subsection{Comparison of results}
In Sections \ref{lumhon}, \ref{bayes} and \ref{evo}
we derived the quasar halo occupation number using a few different methods.
The corresponding outcomes are fully consistent with each other.
In all cases, we find that bright quasars are hosted by massive
haloes with $M\simgt 10^{12} M_\odot$. 
For larger halo masses,
the shape of the halo occupation number
is not well constrained by the observational data and a wide range
of possibilities is allowed.
However, independently of the model details,
we find that quasar hosts have characteristic masses
of a few $\times 10^{13} M_\odot$.
This is the key result of this analysis which
strongly constrains quasar formation models.
For instance,
by coupling hydrodynamical simulations of galaxy formation with 
simple recipes for AGN activation, Di Matteo et al. (2003) 
recently concluded that quasar hosts at $z\sim 2$ have typical
masses of $\sim 4\times 10^{12} M_\odot$. The corresponding
clustering amplitude ($b\sim 1.6$ at $z=1.89$) 
is too low to match our measures ($b=3.9\pm 0.3$ at $z_{\rm eff}=1.89$),
thus suggesting that some revision of the model is probably required.

On the other hand, our findings are in good 
agreement with the typical mass of haloes hosting
local radio galaxies (Magliocchetti et al. 2004).
This further stengthens the connection between active galactic nuclei which 
exhibit different observational properties.

\subsection{Control of systematics}
A number of assumptions have been used in the present study.
We discuss here how possible sources of systematic errors
might affect our results.

All our analysis is developed within a specified cosmological framework 
based on the CDM paradigm.
Modifying the cosmological parameters within the ranges allowed
by recent CMB studies (e.g. Tegmark et al. 2004)
induces minor changes in our conclusions. 
Results similar to those presented here are also obtained by slightly 
altering the power spectrum of density fluctuations.
For instance, neglecting the presence
of baryons (i.e. modifying the shape parameter of the linear power
spectrum from 0.16 to 0.21) increases the bias parameters of our sub-samples
by $\sim 7$ per cent. In consequence, 
our best-fitting values for $\alpha$ increase by 0.2-0.3. 
At the same time, for a given $\alpha$, the best-fitting
values for $\log_{10}M_0$ and $\log_{10} N_0$ increase by 0.2-0.4.

The normalisation of the linear power spectrum of density fluctuations
is still very controversial: estimates of $\sigma_8$ from weak-lensing
and cluster abundances range between 0.7 and 1 (see e.g. 
Table 4 in Tegmark et al. 2004 for a list of the most recent determinations).
In Table \ref{Tab:s8}, we use our entire sample of quasars
to show how the best-fitting parameters of the halo model 
change with $\sigma_8$.
For simplicity, we just consider models with $\alpha=0$ and $\alpha=1$.
Note that, while 
the estimated bias parameter and $\sigma_8$ are inversely proportional,
the best-fitting parameters of the halo model depend only slightly
on the assumed value for $\sigma_8$ 
(compared with their statistical uncertainty). 

Our analysis relies on a set of fitting functions calibrated
against numerical simulations. These have been used to compute
the mass function and bias parameter of dark matter haloes
and the non-linear power spectrum of density fluctuations.
Considering all the uncertainties, 
we estimate that, on the scales considered here,
the accuracy of the resulting correlation function is 
of the order of 10-20 per cent.
This is still smaller than the statistical error associated
with the observed correlation function. In consequence, we do not
expect our results to be significantly affected by this source
of systematic errors.

We used the most recent observational determinations of the Eddington ratio
and of the correlation between $M_{\rm bh}$ and $v_{\rm c}$
to estimate the mass function of quasar host haloes.
What is the sensitivity of our results to these assumptions?
Assuming that all
high-redshift quasars shine at the Eddington luminosity 
(which is a bit extreme but certainly plausible)
would decrease our estimates for $M_{\rm bh}$ by a factor 2-3
and the mass of the host-haloes by a factor of 3-5.
The best-fitting solutions for $N(M)$ would then 
correspond to values for $\alpha$ which are slightly larger than those
presented in Section \ref{sol}.

A large fraction of quasar-host galaxies are
morphologically disturbed or interacting.
This suggests that 
efficient black-hole fueling is triggered by galaxy encounters 
involving at least one gas rich object.
Based on this, Kauffmann \& Haehnelt (2000) developed a merger-based 
prescription of AGN activation. 
In our analysis,
we never distinguish between merging and non-merging haloes.
Can this bias our results?
Previous studies have shown that, at $z\sim 2$,
merging and randomly selected haloes of the same mass have the
same clustering properties (Kauffmann \& Haehnelt 2002; Percival et al. 2003). 
This implies that our results are valid also in the merger-driven 
scenario for AGNs.
However, if quasars are indeed found only in merging haloes, our estimates
for $t_{\rm Q}$ should be revised upwards by a factor of $f^{-1}_{\rm mer}$
with $f_{\rm mer}$ the fraction of merging haloes.

\begin{table}
\begin{center}
\caption{Dependence of the best-fitting bias and 
halo-model parameters on $\sigma_8$. The first set of data refers
to models with $\alpha=0$ and the second to $\alpha=1$.
The entire quasar sample
($0.8<z<2.1$) is considered here.
\label{Tab:s8}} 
\begin{tabular}{cccccc}
\hline
\hline
$\sigma_8$ & $b$ & 
$\log_{10}\displaystyle{\frac{M_0}{M_\odot}}$ & $\log_{10} N_0$ & $\log_{10}\displaystyle{\frac{M_0}{M_\odot}}$ & $\log_{10} N_0$\\
& & \multicolumn{2}{c}{$\alpha=0$} & \multicolumn{2}{c}{$\alpha=1$}
\\
\hline
0.7 & $2.76\pm 0.23$ & $12.5^{+0.1}_{-0.2}$ & $-1.8^{+0.2}_{-0.3}$
& $12.0^{+0.2}_{-0.4}$ & $-2.9^{+0.3}_{-0.5}$
\vspace*{0.1cm}\\
0.8 & $2.42\pm 0.20$ & $12.6^{+0.1}_{-0.2}$ & $-1.8^{+0.2}_{-0.3}$
&$12.0^{+0.2}_{-0.4}$ & $-3.1^{+0.3}_{-0.5}$
\vspace*{0.1cm}\\
0.9 & $2.15\pm 0.18$ & $12.6^{+0.1}_{-0.2}$ & $-1.9^{+0.2}_{-0.3}$
&$11.9^{+0.2}_{-0.4}$ & $-3.3^{+0.3}_{-0.5}$
\vspace*{0.1cm}\\
1.0 & $1.91\pm 0.16$ & $12.7^{+0.1}_{-0.2}$ & $-1.9^{+0.2}_{-0.3}$
&$11.7^{+0.2}_{-0.4}$ & $-3.6^{+0.4}_{-0.6}$
\vspace*{0.1cm}\\
\hline
\end{tabular}
\end{center}
\end{table}

\begin{figure*}
\centerline{\epsfig{figure=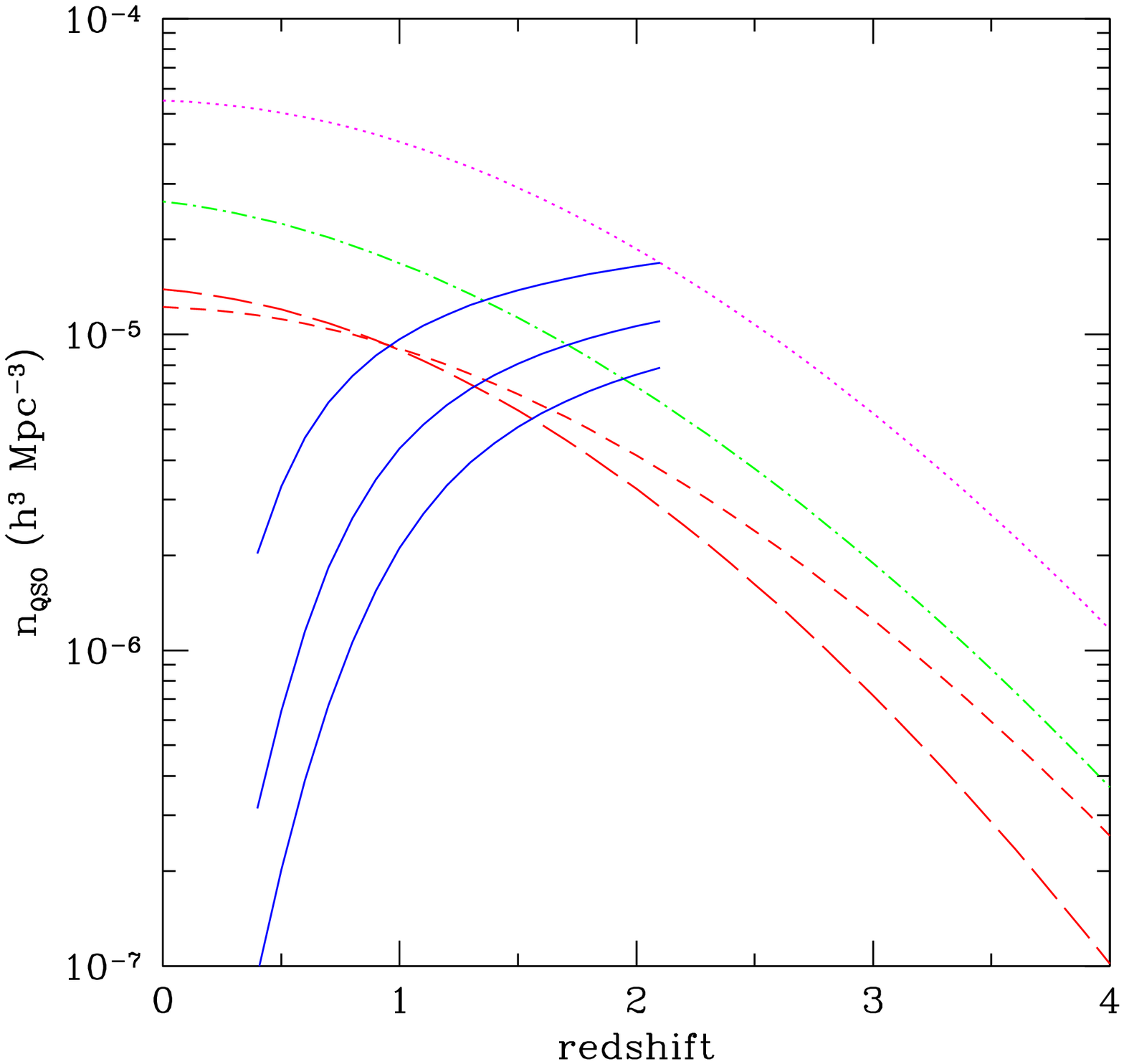,height=8cm}
\epsfig{figure=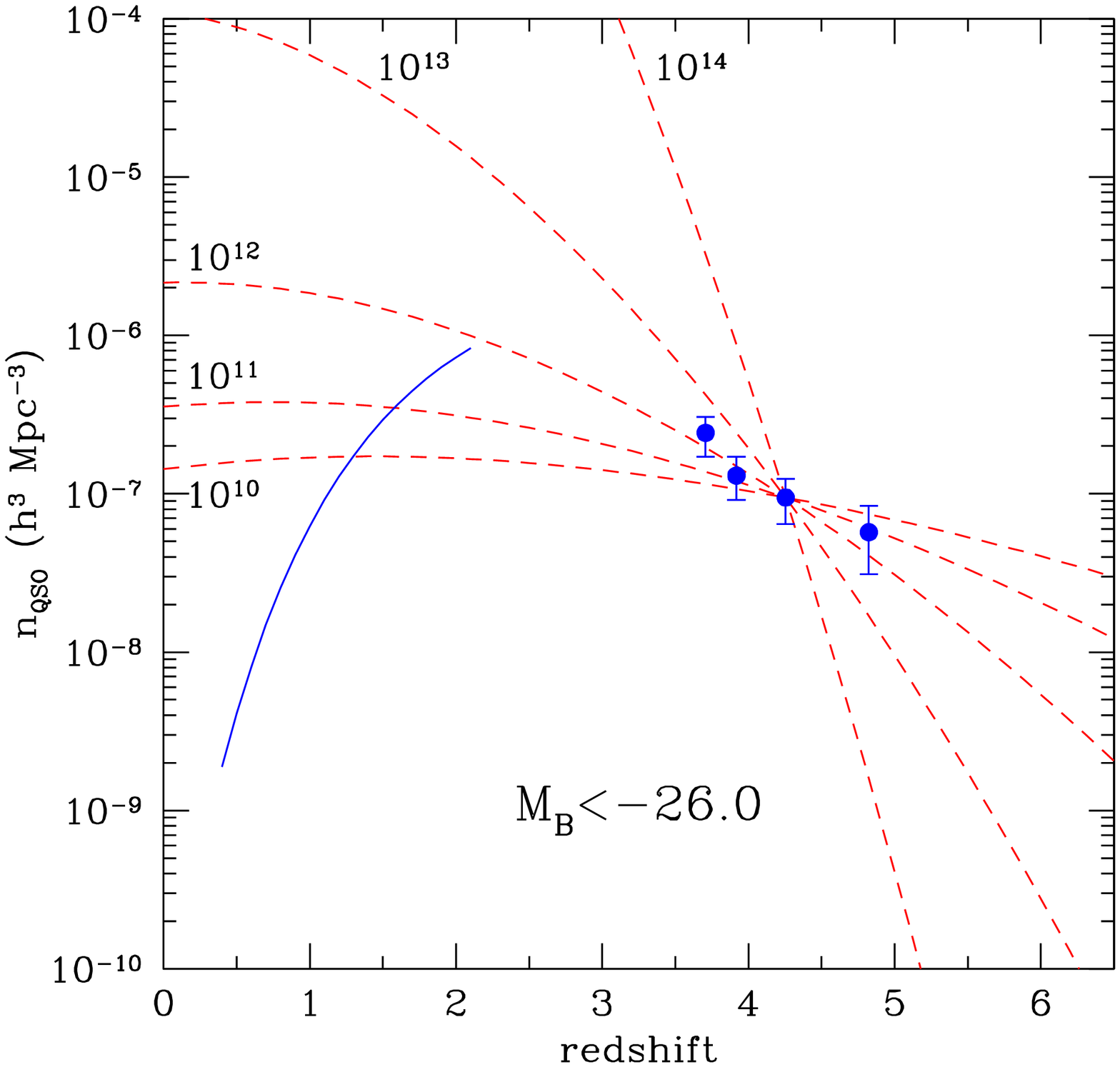,height=8cm}}
\caption{Number-density evolution of optically bright quasars.
{\it Left panel:}
The solid lines are obtained using the best-fitting luminosity function 
from the 2QZ (Croom et al. 2004). From top to bottom they refer to
$M_{b_{\rm J}}<-22.5,-23.6,-24.1$ (which correspond to the faintest
objects in our sub-samples).
The remaining lines show the evolution of $n_{\rm QSO}$ corresponding
to a fixed halo occupation number. 
Two best-fitting models for $N(M)$ at $z_{\rm eff}=1.06$
are represented with dashed lines:
namely, the Prior B solution discussed in Section \ref{sol} (short-dashed)
and the non-evolving model presented in Section \ref{evo} (long-dashed).
The dot-dashed line shows the best-fitting (Prior B) model for our full sample.
The dotted line is obtained by renormalising the short-dashed line
so to fit the 2QZ data at $z=2.1$ (which corresponds to $\log_{10}N_0=-1.65$).
{\it Right panel:}
%
Datapoints with errorbars show the high-redshift results 
from the SDSS Quasar Survey (Fan et al. 2001).
The solid lines are obtained using the best-fitting luminosity function 
from the 2QZ (Croom et al. 2004). 
The dashed lines refer to halo occupation models of the form
$N(M)=N_0 \cdot \Theta(M-M_0)$. 
The adopted values for $M_0$ are indicated in the 
figure and $N_0$ is fixed so to match the observed quasar density at $z\sim 4$.
\label{density}}
\end{figure*}

%
%
%

\subsection{Number-density evolution: implications for high-redshift quasars}

It is interesting to study how the number density of quasars 
with a given halo occupation number evolves.
This is shown, for different halo models,
in the left panel of Fig.~\ref{density}.
In all cases, the number density rapidly drops with redshift
as a consequence of the hierarchical assembly of dark-matter haloes
(see also Efstathiou \& Rees 1988).
On the other hand, by integrating the 2QZ luminosity function 
(Croom et al. 2004) above a given threshold value, 
one finds that, between $0.4<z<2.1$,
$n_{\rm QSO}$ increases with lookback time (see Fig.~\ref{density}).
This is clearly seen also in Table \ref{Tab:data}:
both our low-redshift sub-sample and our full sample
roughly correspond to $M_{b_{\rm j}}<-22.5$ but
the quasar number density at $z_{\rm eff}=1.47$ is a factor 1.3 higher
than at $z_{\rm eff}=1.06$.
We then conclude that, at $z<2$ and for a given luminosity threshold, 
the quasar halo occupation number cannot stay constant with time:
at least its overall normalisation, $N_0$, 
(and the corresponding quasar lifetime) has to increase with $z$.
This is probably due to the fast depletion of the gas available for
accretion onto supermassive black holes during the late stages of
galaxy and group formation (e.g. Cavaliere \& Vittorini 2000).
Once again it is important to stress the different nature of
the halo-model parameters. Basically, while $M_0$ determines which haloes
are capable of hosting supermassive black holes,
$N_0$ and $\alpha$ fix the overall normalisation and the scaling
of the halo occupation number with the halo mass. 
These two parameters are probably 
more influenced (with respect to $M_0$)
by the local physics which determines the efficiency of gas accretion.

It is reasonable to expect that a fixed halo occupation
number might accurately describe the quasar density evolution at higher
redshifts
when gas is ubiquitously available within massive dark matter haloes.
In fact, there is a consensus that
the comoving number density of optically selected quasars peaks at $z\sim 2-3$
and drops rapidly at higher redshifts.
\footnote{
A number of factors (namely, observational incompletenesses, uncertainties
in the K-corrections and the possibility that a large fraction
of quasars is not detectable in the optical band due to dust extinction)  
could generate a spurious drop but it is widely believed that at least
part of the observed decrease is real 
(see e.g. the discussion in Fan et al. 2001).}
This is naturally explained by CDM models 
where galaxies form at relatively late times (Efstathiou \& Rees 1988).
In the right panel of Fig.~\ref{density}, 
we compare the density evolution predicted
by the halo model with observational data
from the 2QZ and the SDSS Quasar Survey (Fan et al. 2001).
For simplicity, we set $\alpha=0$ 
and we assume that the function $N(M)$ does not evolve with time.
We find that models with $M_0\simgt 10^{12} M_\odot$ are consistent
with the observed number density of quasars with $M_{\rm B}<26$ 
in the redshift interval $2\simlt z \simlt 5$.
On the other hand, as discussed above, a fixed $N(M)$ cannot match the 
data at $z<2$. 
Similar results are also obtained for a brighter sample 
of $i$-dropout objects detected at $z\sim 6$ (Fan et al. 2004).

Knowledge of the number density evolution, however, 
does not provide enough information to determine
the nature of high-redshift quasars. In fact, different 
halo occupation models
which are compatible with the observed evolution of $n_{\rm QSO}$ 
correspond to wildly discrepant quasar characteristics.
For instance, at $z=4$,
a model with $\alpha=0$ and $M_0=10^{12} M_\odot$
corresponds to a correlation length of  $6.7$ \Mpc \ ($b=5.0$)
and to a mean host-halo mass 
of $\langle M \rangle= 1.8 \times 10^{12} M_\odot$.
In this case, the observed number density implies that $N_0=3.8\times 10^{-4}$
and $t_{\rm Q}=2.5 \times 10^5$ yr.
Thus, assuming that equation (\ref{Mv}) still holds at $z=4$, 
from the estimated $\langle M \rangle$, 
one derives $M_{\rm bh}=1.2 \times 10^8 M_\odot$
which implies a super-Eddington accretion rate with $\eta=2.8$. 
On the other hand, the corresponding results for 
a model with $\alpha=0$ and $M_0=10^{13} M_\odot$ are:
$N_0=0.18$, $r_0=14$ \Mpc \ ($b=9.2$),
$\langle M \rangle=1.4 \times 10^{13} M_\odot$, $t_{\rm Q}=1.3
\times 10^8$ yr, $M_{\rm bh}=2.1 \times 10^9 M_\odot$
and $\eta=0.16$.
This clearly shows that future clustering measurements 
(hopefully combined with information on $t_{\rm Q}$)
will be crucial
to understanding the physical properties of high-redshift quasars.
The low abundance of optically bright objects, however, poses
enormous difficulties for this kind of studies. 

\section{Summary}
\label{summary}
We have used a flux limited sample of $\sim 14,000$ 2QZ quasars
with $M_{b_{\rm J}}<-22.5$ 
to study the quasar clustering properties in the redshift range $0.8<z<2.1$. 
Our main results are summarized as follows.

(i) For spatial separations between 1 and 20 $h^{-1}$ Mpc,
the correlation function for our whole quasar sample 
(corresponding to an effective redshift $z_{\rm eff}=1.47$)  
is well approximated by a power law with
slope $\gamma=1.53\pm 0.20$ and comoving
correlation length $r_0=4.8^{+0.9}_{-1.5}$ \Mpc.

(ii) Splitting the sample into three redshift ranges, we find evidence
for an increase of the clustering amplitude with lookback time.
The correlation function for quasars at $1.7<z<2.1$ ($z_{\rm eff}=1.89$) 
is nearly a factor of 2 higher with respect to the whole sample 
($z_{\rm eff}=1.47$).
Since flux-limited surveys tend to select intrinsically brighter objects at 
higher-redshifts,
it is not possible to tell, however, whether this effect is due to 
real evolution of the quasar population or to luminosity dependent clustering.
We will further address this issue in a future paper.

(iii)
For all the sub-samples, 
the correlation function is well approximated by a power law.
The best-fitting parameters, which are strongly covariant
(see Fig. \ref{Fig:data}), 
range between $-2.0\simlt\gamma\simlt-1.5$ 
and $4\simlt r_0/{h^{-1} {\rm Mpc}}\simlt 8$ (see Table \ref{Tab:PL}).
Within the statistical uncertainties, data in different redshift bins
can be anyway described by the same value of $\gamma$.
Assuming that the slope of the correlation function 
does not change with redshift, 
evolution of the correlation length 
is detected at the $3.6\, \sigma$ confidence level.

(iv)
Within the framework of concordance cosmology, high-redshift quasars
are more biased tracers of the mass distribution than their low-redshift
counterparts. 
The observed quasar-to-mass bias parameter is consistent with being 
scale-independent for the separations probed by our analysis.
Assuming $\sigma_8=0.8$, we obtain
$b=2.42^{+0.20}_{-0.21}$ for the whole quasar sample.
On the other hand, we find $b=1.80^{+0.20}_{-0.24}$
for $0.8<z<1.3$, $b=2.62^{+0.18}_{-0.19}$ for $1.3<z<1.7$ and
$b=3.86^{+0.32}_{-0.35}$ for $1.7<z<2.1$. 
In hierarchical models for structure formation, the bias parameter of a 
population of tracers can
be readily linked to the mass of their host dark-matter haloes.
At a given $z$,
values of $b$ which are significantly larger than unity correspond to
haloes with $M\gg M_*(z)$ where $M_*(z)$ denotes the characteristic mass of
haloes which are forming at that epoch out of $1\,\sigma$ density fluctuations.
The bias parameters of our sub-samples then suggests that 2QZ quasars
are hosted by rare haloes with $M\sim 10^{13} M_\odot$.

(v)
Using the halo model, we find
that the observed quasar number density and clustering amplitude are consistent
with a picture where: 
(a) quasars form in haloes with $M>10^{12} M_\odot$;
(b) the characteristic 
mass of their host haloes is a few $\times 10^{13} M_\odot$.
This result 
is independent of the detailed form of the halo occupation number
and hence it can be used to constrain models of quasar formation.

(vi)
Our best-fitting models at $z_{\rm eff}=1.06$ suggest that 
$N(M)\propto M^{0.4-0.5}$ for $M>10^{12} M_\odot$ and rapidly drops to zero 
for smaller values of $M$.  
For higher redshifts, $N(M)$ tends to increase more rapidly with the halo mass.
For instance, at  $z_{\rm eff}=1.89$, $N(M)\propto M^{1-1.5}$ for 
$M>10^{13} M_\odot$.
It is worth stressing, however, that the
data are also consistent with a non-evolving functional form for $N(M)$ 
where quasars
reside in haloes more massive than $5\times 10^{12} M_\odot$ 
and where 
the halo occupation number has a very weak dependence on the halo mass.

(vii) The mean number of quasars per halo is always much smaller than one.
Systematic searches for close pairs
are needed to understand whether 2 active quasars can be hosted by a single
halo.

(viii) The observed clustering evolution is consistent with assuming
that the locally observed correlation between black-hole mass and
host-galaxy circular velocity (Ferrarese 2002; Baes et al. 2003)
is still valid at $z>1$.

(ix) 
The fraction of potential host-haloes which indeed harbour a bright quasar 
increases from $\simlt 1$ per cent at $z_{\rm eff}=1.06$ to $5-10$ per cent
at $z_{\rm eff}=1.89$. From this, we infer 
that the characteristic quasar lifetime $t_{\rm Q}$
increases with redshift (and/or with optical luminosity), ranging
from a few $\times 10^7$ yr at $z\sim 1$ to $\sim 10^8$ yr at $z\sim 2$.
This is in good agreement with studies of the proximity effect
(e.g. Jakobsen 2003 and references therein).

(x) For $z<2$, the halo occupation number of quasars which are above
a given absolute luminosity threshold cannot stay constant with time.
In order to match the observed number density evolution, at least
its overall normalization $N_0$ (and thus the corresponding $t_{\rm Q}$)
has to increase with $z$. This probably reflects the fast depletion
of the gas available for accretion onto supermassive black holes
during the process of galaxy formation.

In brief,
this paper presents state-of-the-art measurements of quasar clustering
and establishes an accurate benchmark for quasar-formation models.
Future results from the SDSS quasar survey will provide an independent
verification of our results and, 
thanks to the different quasar-selection criteria, 
will extend them to even higher redshifts.

\section*{ACKNOWLEDGMENTS}

CP and PN thank Simon Lilly for useful suggestions and comments.
MM acknowledges 
Gianfranco De Zotti and Luigi Danese for many clarifying discussions.
PN thanks Phil Outram and Tom Shanks for stimulating discussions about
the 2QZ.
We are grateful to the anonymous referee for valuable suggestions
that improved the presentation of our results.
CP and PN are supported by the Zwicky Prize fellowship program at 
ETH-Z\"urich. 
The 2dF QSO Redshift Survey (2QZ) was compiled by the 2QZ survey team
from observations made with the 2-degree Field on the Anglo-Australian
Telescope.

\appendix

\section{The Mass distribution of quasar host haloes}
\label{honlum}

In this section, we use a few observational results
to derive the conditional probabiliy distribution of the host-halo mass, $M$,
for a quasar with given absolute magnitude $M_{b_{\rm J}}$.

\subsection{From photographic to Johnson $B$ magnitudes}
We start from converting $b_{\rm J}$ fluxes into standard $B$ magnitudes.
In general, 
$B\simeq b_{\rm J}+0.3\,(B-V)$ (Blair \& Gilmore 1982; Colless et al. 2001),
and the rest-frame color index for quasars is $B-V\simeq 0.22$ 
(Cristiani \& Vio 1990).
In what follows, we then assume 
\begin{equation}
M_{\rm B}=M_{b_{\rm J}}+0.07\;,
\label{bj2b}
\end{equation}
which is in good agreement with Brotherton et al. (2001). 
Note that the amplitude of the correction is comparable with the statistical
error which affects the magnitude determination in the 2dFGRS
(Colless et al. 2001; Norberg et al. 2002b) which uses the same UKST
photographic plates as the 2QZ. 
It is then reasonable to expect that also quasar photometry in the 2QZ 
is affected by typical errors of $\sim 0.1$ magnitudes
(e.g. Corbett et al. 2003).

\subsection{Bolometric corrections}
\label{bcorr}
In order to use equation (\ref{Mbh}) to infer the mass of the black holes
which power 2QZ quasars,
we need to convert their absolute $B$ magnitudes into bolometric luminosities.
Bolometric corrections for a sample of X-ray selected quasars lying at $z<1$
have been derived in a seminal paper by Elvis et al. (1994). 
Since observations show that quasar spectra do not evolve with $z$
(e.g. Bechtold et al. 1994), it is common practice to apply these corrections
also to high-redshift quasars.
It has been recently pointed out, however, that the bolometric corrections
by Elvis et al. (1994)  
are seriously affected by systematics and should be revised downwards
(e.g. Fabian \& Iwasawa 1999; Elvis, Risaliti \& Zamorani 2002).
For this reason we use here the results
by McLure \& Dunlop (2004) who,
adopting the revised template spectrum by Elvis et al. (2002),
estimated the bolometric corrections
for 1136 quasars at $0.5<z<0.8$ extracted from the 
SDSS. 
Corrections from the B band have then been computed using a 
subsample of 372 objects common to the the 2dF and SDSS surveys.
When combined with equation (\ref{bj2b}), their best fitting relation gives
\begin{equation}
\log_{10} \left(\frac{L_{\rm bol}}{10^{46}\,{\rm erg\,s}^{-1}}\right)=
0.21-0.38\,(M_{b_{\rm J}}+25)
\label{BC_2}
\end{equation}
and the corresponding rms variation at fixed $M_{b_{\rm J}}$ is 0.14.
Within the quoted uncertainties, this is perfectly consistent with the
recent results by Marconi et al. (2004).~\footnote{The bolometric corrections by Marconi et al. (2004) 
are roughly 2/3 of those by Elvis et al. (1994) and
correspond to
$\log_{10} \left(\frac{L_{\rm bol}}{10^{46}\,{\rm erg\,s}^{-1}}\right)=
0.37-0.40\,(M_{b_{\rm J}}+25)$
with a scatter at fixed $M_{b_{\rm J}}$ of 0.3.}

\subsection{The Eddington ratio and the distribution of black-hole masses}
Observational estimates of the Eddington ratio, $\eta$, require:
(i) using some dynamical tracer to determine the black-hole mass
(and thus the Eddington luminosity);
(ii) measuring the quasar luminosity in a given band, $L_i$;
(iii) applying the corresponding bolometric correction, 
$\beta_i^{\rm bol}$;
(iv) calculating $\eta\propto \beta_i^{\rm bol}\, L_i/M_{\rm bh}$.
Given this complexity, measurements of $\eta$ 
are rather uncertain and sensitive to a number of sources of systematic 
errors. Recent determinations, however, tend to lie in the same ballpark and
suggest that $\eta$ mildly increases with $z$ (Dunlop et al. 2003; McLure \&
Dunlop 2004).
For consistency with Section \ref{bcorr}, we use here the results by
McLure \& Dunlop (2004) who
combined virial estimates of black hole masses 
with new bolometric corrections 
to infer the Eddington ratio
for a large sample of quasars from the SDSS.
From their results we infer that, for $0.8<z<2.1$, the
probability density function for $\log_{10} \eta$ is well 
approximated by a Gaussian distribution with mean $0.21 z-0.80$
and variance $\sim 0.3$.
This corresponds to $\langle\eta\rangle=10^{0.21 z-0.65}$.
These results are also supported by other indirect determinations of 
$\eta$.
By requiring the mass function of relic black holes (as inferred from the
X-ray background) to match its local counterpart,
Marconi et al. (2004) found that $0.1\simlt \eta \simlt 1.7$ 
(with a preferred value of $\eta \sim 0.5$). 
Once accounted for the different bolometric corrections,
these values are in extremely good agreement with the results from 
McLure \& Dunlop (2004).
Similarly, Yu \& Tremaine (2002) showed that the local mass density in black 
holes is consistent with the integrated luminosity density of quasars if 
they accreted mass nearly at the Eddington rate at redshifts $z\simgt 2$.

\subsection{The distribution of black-hole masses}
\label{bhm}
The conditional probability distribution of $\log_{10} M_{\rm bh}/M_\odot$ 
for a given $L_{\rm bol}$ is thus obtained by combining equation (\ref{Mbh})
with the observationally determined distribution of $\eta$:
\begin{equation}
P(\log_{10} \frac{M_{\rm bh}}{M_\odot}|L_{\rm bol})=0.73\,
\exp\left[-\frac{(\log_{10} \frac{M_{\rm bh}}{M_\odot}
-f)^2}{0.6}\right]
\label{Pbh}
\end{equation}
with $f=8.70-0.21\,z+\log_{10} (L_{\rm bol}/10^{46} {\rm erg \, s}^{-1})$.
For consistency, in order to estimate the black-hole mass associated with
a quasar of given absolute magnitude $M_{b_{\rm J}}$,
we combine equations (\ref{BC_2}) (including its associated scatter) 
and (\ref{Pbh}) which have been derived from the same dataset.
This implies that a quasar with $M_{b_{\rm J}}=-25$ corresponds to a mean
black-hole mass of $5.13\times10^8\,M_\odot$ at $z=1$ and
of  $3.16\times10^8\,M_\odot$ at $z=2$.
Note that for the typical redshift and magnitude ranges spun 
by our sample, 
$6\times 10^7\,M_\odot
\simlt \langle M_{\rm bh}|M_{b_{\rm J}}\rangle \simlt 3\times 10^9\,M_\odot$.
This interval is consistent with the masses inferred from dynamical
measures in the local Universe (e.g. Tremaine et al. 2002 and references
therein) and from the analysis of  
emission linewidths in the 2QZ (Corbett et al. 2003).

\subsection{The mass of host haloes}
\label{halomass}
Taking a step further, we can estimate the probability distribution that
a quasar of a given luminosity is hosted by a dark-matter halo of mass
$M$.
Following Ferrarese (2002; see also Baes et al. 2003),
this is obtained by assuming that a statistically significant correlation 
links $M_{\rm bh}$ and $M$. 

Black-hole masses are found to be tightly correlated with  the velocity 
dispersion of their host spheroid, $\sigma_{\rm sph}$ 
(Ferrarese \& Merritt 2000; Gebhardt et al. 2000).
The most recent determination considers $\sim 30$ galaxies with secure
detections of supermassive black holes (Tremaine et al. 2002).
Observations also provide evidence 
for a correlation between
$\sigma_{\rm sph}$ and the circular velocity 
in the flat part of the rotation curve of the host galaxy
(Ferrarese 2002; Baes et al. 2003). 
By combining the $M_{\rm bh}-\sigma_{\rm sph}$ 
and the $\sigma_{\rm sph}-v_c$ relations, Baes et al. (2003) find
\begin{equation}
\frac{M_{\rm bh}}{M_\odot}=
10^{7.24\pm 0.17} \left(
\frac{v_{\rm c}}{200\, {\rm km\,\, s^{-1}}} \right)^
{4.21\pm 0.60}
\;.
\label{Mv}
\end{equation}
This purely observational relation
can be used to link $M_{\rm bh}$ with the mass of the host halo. 
In order to do this, however, one needs to express $v_{\rm c}$ in terms
of $M$ which is a formidable task.
As a first order approximation one can assume an
equilibrium configuration for the dark-matter density profile in haloes.
Both the singular isothermal sphere and models derived 
from numerical simulations (e.g. Navarro et al. 1997)
provide good starting points.
However, detailed modelling of the rotation curve requires accounting for
the distribution and physics of baryons (e.g. Mo, Mao \& White 1998).
In fact, the gas contribution can be dominant in the innermost
regions of galaxies. Moreover,
the condensation towards the centre of the dissipative material  
redistributes, through gravity, the collisionless matter.

For simplicity, we consider equilibrium profiles which only
contain dark matter and we account for the presence of baryons
in an approximate way.
The circular velocity at the virial radius of each halo is
\begin{equation}
\frac{v_{\rm vir}}{159.4\,{\rm km\, \,s^{-1}}}=
\left( \frac{ M }{ 10^{12}\,h^{-1}\,M_\odot} \right)
 ^{1/3}
\left(\frac{E_z^2\,\Delta_z}{18\,\pi^2} \right)^{1/6}\;,
\label{vcirc}
\end{equation}
where $E_z^2=\Omega_0\,(1+z)^3+\Omega_\Lambda$, 
$\Delta_z$ is the ratio between the mean density of the halo and
the critical density of the Universe (both evaluated at redshift $z$).
For a spherical collapse, this function can be approximated as
$\Delta\simeq 18\pi^2+82x-39x^2$ with $x=\Omega_{\rm m}(z)-1$ and
$\Omega_{\rm m}(z)=\Omega_0\,(1+z)^3/E_z^2$ (Bryan \& Norman 1998).

A truncated singular isothermal sphere has a constant circular velocity profile
$v_{\rm c}=v_{\rm vir}$,
while for an NFW density profile 
\begin{equation}
\frac{v_{\rm c}({\cal R})}{v_{\rm vir}}=
\left[\frac{1}{{\cal R}}\frac{F({\cal C} \cdot {\cal R})}
{F({\cal C})}\right]^{1/2}\;,
\end{equation}
where ${\cal R}=r/r_{\rm vir}$, $F(x)=\ln (1+x)-x/(1+x)$ and ${\cal C}$ is the
concentration parameter of the halo.
In this case, the circular velocity vanishes when ${\cal R}\to 0$, 
reaches a maximum at 
${\cal R}\simeq2.16/{\cal C}$ and matches the virial velocity
at ${\cal R}=1$. 
Two questions naturally arise: 
{\it i)} What is the value of ${\cal R}$ which corresponds
to the observed circular velocities? 
{\it ii)} What is the contribution of the baryons at this radius?
These are the main uncertainties of our analysis.

For galaxies with HI rotation curves, $v_{\rm c}$ is typically measured 
at a few tens of kpc from the centre, well beyond the optical radius
(a few kpc).
On the other hand, the present-day 
virial radius of a halo with $M=10^{13} M_\odot$
is $r_{\rm vir}=0.56$ Mpc. 
In other words,
the largest scales sampled by rotation-curve measurements 
are nearly a factor of 10 smaller 
than the virial radius.
Using galaxy-galaxy lensing data from the SDSS,
Seljak (2002) has shown that, for galaxies above $L_*$, 
$v_{\rm c}$ decreases significantly
from the optical radius of a galaxy to the virial radius of its host halo.
This result is independent of the morphological type and is probably
suggesting that baryons contribute significantly to the
circular velocity at the optical radius and that density profiles for
the dark matter are highly concentrated (as expected in CDM models at $z=0$). 
Seljak (2002) also found a clear trend for
the ratio $\psi=v_{\rm c}/v_{\rm vir}$ with halo mass.
Typical values are:
$\psi\sim 1.8$ for $M\sim 3\times 10^{11} M_\odot$,
$\psi\simeq 1.4\pm 0.2$ for $M\sim 10^{13} M_\odot$ 
and $\psi<1$ for cluster masses.
This is in good agreement with the predictions of CDM models, since the dark 
matter concentration is expected to decrease with the halo mass 
and the baryonic contribution is expected to become less and less important.

Assuming that the observed $v_{\rm c}$ corresponds to the
maximum value of the rotational velocity profile in an NFW halo 
tends to underestimate  Seljak's results. 
Using equations (9) and (13) in Bullock et al. (2001), we find that, at $z=0$,  this assumption corresponds to $\psi=1.3$ (${\cal C}\sim 14$)
for $M\sim 3\times 10^{11} M_\odot$ and $\psi=1.2$ (${\cal C}\sim 9$) 
for $M\sim 10^{13} M_\odot$. Anyway, these results show
the correct trend with the halo mass:
smaller, more concentrated haloes are associated with larger values for $\psi$.
It is worth noticing, however, that the candidate hosts of our quasars
(haloes with $M\sim 10^{13} M_\odot$ at $0.8<z<2.1$)
are expected to be much less concentrated (${\cal C}\sim 3-5$) than 
their present-day counterparts. 
In this case, the maximum value of the rotational velocity is only 
2-15 per cent higher than $v_{\rm vir}$. This motivates the choice
$\psi\simeq 1$ as a viable alternative to the low-redshift results
by Seljak (2002).

We have now collected all the elements necessary to estimate 
the conditional probability distribution of the host-halo mass, $M$, 
for a quasar with given absolute magnitude $M_{b_{\rm J}}$: 
${\cal P}(M| M_{b_{\rm J}})$.
In brief: 
{\it i)} We assume that equations (\ref{BC_2}), (\ref{Pbh})
and (\ref{Mv}), which have been determined at lower redshifts, are still valid 
for the host galaxies of our 2QZ quasars at $0.8<z<2.1$;
Their combination (including the scatter in each of them) is used to determine
the probability distribution of $v_{\rm c}$; 
{\it ii)} We then convert circular velocities into halo masses by
selecting a value of $\psi$ and using equation (\ref{vcirc}).
To match the results by Seljak (2002),
we assume a Gaussian distribution for $\psi$ with mean value
of 1.4 and scatter of 0.2 (case A); alternatively, based on the estimated
low concentration of high-redshift haloes, we assume that $\psi=1$ (case B).

\section{The halo occupation distribution from semi-analytic models of galaxy formation}

We use here semi-analytic models of galaxy formation to get an insight
into the problem of choosing a functional form for the first two
moments of the quasar halo occupation distribution.
%

\subsection{The halo occupation number}

\begin{figure}
\centerline{\epsfig{figure=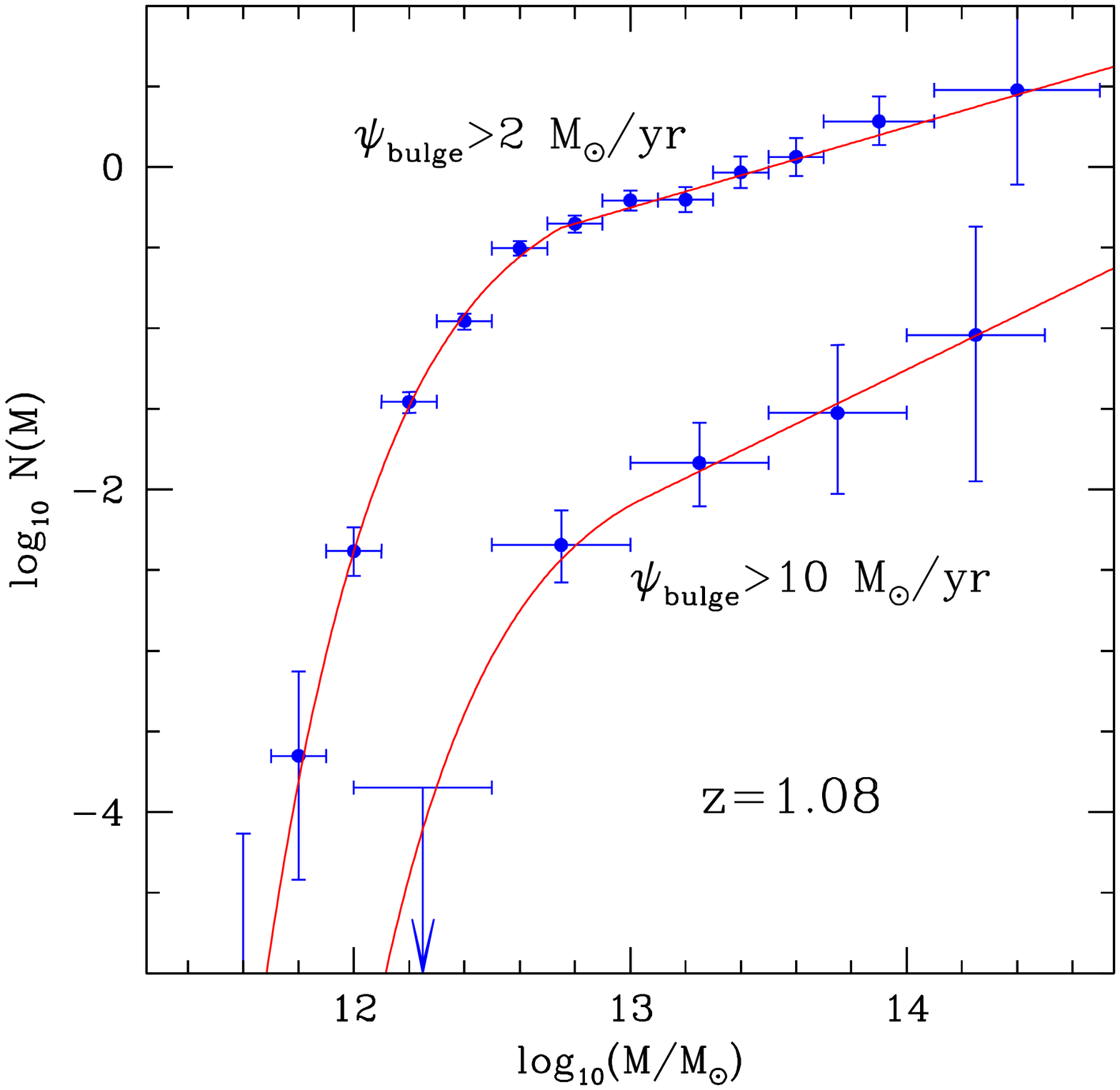,height=8cm}}
\caption{The halo occupation number of galaxies which at $z=1.08$ are
actively forming stars in the bulge as
obtained from semianalytic models of the GalICS collaboration. 
Datapoints correspond to the estimated $N(M)$, 
vertical errorbars mark the associated 
$1\,\sigma$ uncertainties (assuming Poisson statistics for
both the number of galaxies and the number of haloes in a bin),
while the horizontal errorbars denote the size of the mass bins. 
Arrows mark the upper limit for $N(M)$ in the bins where we measure $N=0$.
The continuous lines show a fit to the data obtained by using 
the function in equation (\ref{broken}). 
For
$\psi_{\rm bulge}>10 \,M_\odot\, {\rm yr}^{-1}$, the best-fitting parameters
are $\alpha= 0.85$, 
$M_0=10^{13} M_\odot$
and $N_0= 8\times 10^{-3}$.
On the other hand, 
for $\psi_{\rm bulge}>2 \,M_\odot\, {\rm yr}^{-1}$,
one gets $\alpha= 0.5$, 
$M_0=10^{12.75} M_\odot$
and $N_0=0.42$. 
\label{galics}}
\end{figure}

There is evidence that, at high redshift, quasars are 
associated with star-forming galaxies (e.g. Omont et al. 2001; 
Hutchings et al. 2002).
It is then plausible to expect that the halo-occupation properties
of quasars might share some similarities with those of galaxies which
show active star formation in their nuclear region.
As an example, we derive here the function $N(M)$
from the semi-analytic models of the
GalICS I collaboration (Hatton et al. 2003) 
at $z=1.08$ (roughly corresponding to the median 
value for our low-redshift sample).
%
Results for galaxies with a bulge star formation rate, 
$\psi_{\rm bulge}$, which is larger than $10\, M_\odot\, {\rm yr}^{-1}$
are shown in Fig. \ref{galics}.
The choice of such a threshold for $\psi_{\rm bulge}$ is motivated by
fact that the 
mean density of these objects ($\sim 12 \times 10^{-6} \,h^3\,{\rm Mpc}^{-3}$) 
is comparable with the mean 
density of our low-redshift quasar sample.
In order to improve the statistics, in Fig. \ref{galics} we also show 
the function $N(M)$ for galaxies with 
$\psi_{\rm bulge}>2 \,M_\odot\, {\rm yr}^{-1}$. 
In both cases, the halo occupation number is well approximated by
a power law with a cutoff at small virial masses. 
%
For instance, the function
\begin{equation}
N(M)=N_0 \times
\begin{cases}
\displaystyle{\left(\frac{M}{M_0}\right)^{\alpha}} 
&\text{if $M\geq M_0$} \\
\displaystyle{\exp\left(1-\frac{M_0}{M}\right)}
&\text{if $M<M_0$} \;,
\end{cases}
\label{broken}
\end{equation}
very closely matches the results of the semi-analytical models in Fig. 
\ref{galics}. 
This is in good agreement with equation (\ref{oldhon}) where
a sharp cutoff replaces the exponential decline at small masses.

%

\subsection{The scatter of $P_N(M)$}

\begin{figure}
\centerline{\epsfig{figure=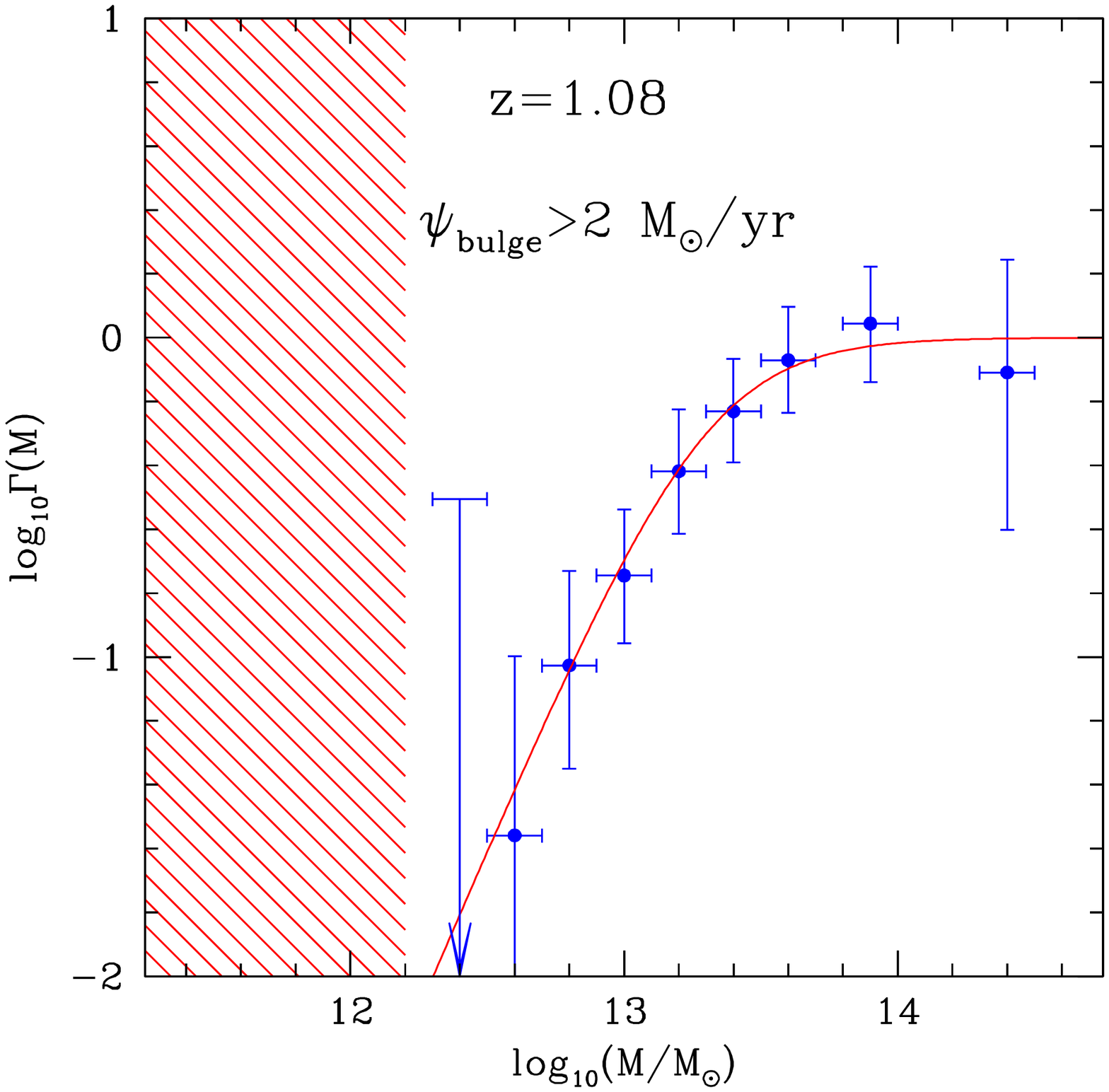,height=8cm}}
\caption{As in Fig. \ref{galics} but for the function $\Gamma(M)$.
The fitting function represented with a continuous line is given
in equation (\ref{gamma}) and corresponds to the best-fitting parameters
$\gamma_{\rm s}=2$ and $M_{\rm s}= 10^{13.3}M_\odot$.
The shaded region indicates the mass range
where the function $\Gamma$ is totally undetermined.
\label{galics2}} 
\end{figure}

In this section, we use
the previously introduced samples of star-forming
galaxies to study the second moment of the halo occupation distribution. 
%
We first note that there is not a single halo in the GalICS I sample which 
contains more
than 1 galaxy with $\psi_{\rm bulge}>10 \,M_\odot\, {\rm yr}^{-1}$ at $z=1.08$.
In other words, the data (within extremely large errorbars) are consistent 
with $\Sigma^2=0$. 
This is clearly an effect of the small number statistics. 
On the other hand, the results 
for $\psi_{\rm bulge}>2 \,M_\odot\, {\rm yr}^{-1}$ are much more significant.  
In this case, the numerical results are well approximated by the function 
\begin{equation}
\Gamma(M)=
\left(\frac{M}{M_{\rm s}}\right)^{\gamma_{\rm s}}\,
\left[1+\left(\frac{M}{M_{\rm s}}\right)^{\gamma_{\rm s}} \right]^{-1}
\label{gamma}
\end{equation}
which scales as a power law for $M\ll M_{\rm s}$ and approaches 1
for $M\gg M_{\rm s}$ (see Fig. \ref{galics2}). 
In agreement with studies of low-redshift galaxies 
(e.g. Sheth \& Diaferio 2001; 
Berlind \& Weinberg 2002; Berlind et al. 2003), 
the scatter of $P_N(M)$ is then 
strongly sub-Poissonian for haloes which, on average, contain
less than 1 object and nearly Poissonian for larger haloes.
%
%
We find that, adopting $N(M_{\rm s})=0.75$ as an operative
definition for $M_{\rm s}$,
equation (\ref{gamma}) with $\gamma_{\rm s}= 2$ 
accurately describes the second moment of the halo occupation distribution
for rare galaxies at $z\simeq 1$. This result does not depend on the
details of the galaxy population considered (star formation rate, colour,
etc.).

\begin{thebibliography}{}
\bibitem[\protect\citename{Adel0}2000]{Adel0}
Adelberger K.~L., 2000,
in Mazure A., Le F\`evre O., Le Brun V., eds., ASP Conf. Ser. Vol. 200,
 Clustering at High Redshift, Astron. Soc. Pac., San Francisco, p. 13 
%
\bibitem[\protect\citename{Adel4}2004]{Adel4}
Adelberger K.~L., 2004, ApJ, in press, astro-ph/0405505
%
\bibitem[\protect\citename{Adel}2003]{Adel}
Adelberger K.~L., Steidel C.~C., Shapley A.~E., Pettini M., 2003, ApJ, 584, 45
%
\bibitem[\protect\citename{Ander}1999]{Ander}
Anderson S.~F., Hogan C.~J., Williams B.~F., Carswell R.~F., 1999, AJ, 117, 56
%
\bibitem[\protect\citename{AC}1992]{AC}
Andreani P., Cristiani S., 1992, ApJ, 398, L13
%
\bibitem[\protect\citename{Aetal}1994]{Aetal}
Andreani P., Cristiani S., Lucchin F., Matarrese S., Moscardini L., 
1994, ApJ, 430, 458
%
\bibitem[\protect\citename{Arn}1999]{Arn}
Arnouts S., Cristiani S., Moscardini L., Matarrese S., Lucchin F., Fontana A.,
Giallongo E., 1999, MNRAS, 310, 54
%
\bibitem[\protect\citename{Baes}2003]{Baes}
Baes M., Buyle P., Hau G.~K.~T., Dejonghe H., 2003, MNRAS, 341, L44
%
\bibitem[\protect\citename{Bah}1997]{Bah}
Bahcall J.~N., Kirhakos S., Saxe D.~H., Schneider D.~P., 1997, ApJ, 479, 642
%
\bibitem[\protect\citename{BDO}1988]{BDO88}
Bajtlik S., Duncan R.~C., Ostriker J.~P., 1988, ApJ, 327, 570
%
\bibitem[\protect\citename{Bech}1994]{Bech}
Bechtold J., Elvis M., Fiore F., Kuhn O., Cutri R.~M., McDowell J.~C.,
Rieke M., Siemiginowska A., Wilkes B.~J., 1994, AJ, 108, 374
%
\bibitem[\protect\citename{Berlind}2002]{berl}
Berlind A.~A., Weinberg D.~H., 2002, ApJ, 575, 587
%
\bibitem[\protect\citename{Berletal}2003]{berletal}
Berlind A.~A. et al., 2003, ApJ, 593, 1 
%
\bibitem[\protect\citename{Blair}1982]{Bla}
Blair M., Gilmore G., PASP, 94, 742
%
\bibitem[\protect\citename{Bro}2001]{Bro}
Brotherton M.~S., Tran H.~D., Becker R.~H., Gregg M.~D., Laurent-Muehleisen S.~A., White R.~L., 2001, ApJ, 546, 775
%
\bibitem[\protect\citename{BryNor}1998]{BryNor}
Bryan G.~L., Norman M.~L., 1998, ApJ, 495, 80
%
\bibitem[\protect\citename{Bullock-2}2002]{bull02}
Bullock J.~S., Wechsler R.~H., Somerville R.~S., 2002, MNRAS, 329, 246 
%
\bibitem[\protect\citename{Bul}2001]{Bul}
Bullock J.~S., Kolatt T.~S., Sigad Y., Somerville R.~S., Kravtsov A.~V., 
Klypin A.~A., Primack J.~R., Dekel A., 2001, MNRAS, 321, 559
%
\bibitem[\protect\citename{Carl}1997]{Carl}
Carlberg R.~G., Cowie L.~L., Songaila A., Hu E.~M., 1997, ApJ, 484, 538
%
\bibitem[\protect\citename{Cate}1998]{Cate}
Catelan P., Lucchin F., Matarrese S. \& Porciani C., 1997, MNRAS, 297, 692 
%
\bibitem[\protect\citename{CV}2000]{CV}
Cavaliere A., Vittorini V., 2000, ApJ, 543, 599
%
\bibitem[\protect\citename{Coil}2001]{coil}
Coil A.~L et al., 2004, ApJ, in press, astro-ph/0305586 
%
\bibitem[\protect\citename{CK }1989]{CK}
Cole S., Kaiser N., 1989, MNRAS, 237, 1127
%
\bibitem[\protect\citename{Coll}2001]{Coll}
Colless M. et al., 2001, MNRAS 328, 1049
%
\bibitem[\protect\citename{Corb}2003]{Corb}
Corbett E.~A., Croom S.~M., Boyle B.~J., Netzer H., Miller L., Outram P.~J., 
Shanks T., Smith R.~J., Rhook K., 2003, MNRAS, 343, 705
%
\bibitem[\protect\citename{Cris90}1990]{cris}
Cristiani S., Vio R., 1990, A\&A, 227, 385
%
\bibitem[\protect\citename{cs}1996]{cs}
Croom S.~M., Shanks T., 1996, MNRAS, 281, 893
%
\bibitem[\protect\citename{Croom01}2001]{croom01}
Croom S.~M., Shanks T., Boyle B.~J., Smith R.~J., Miller L., Loaring N.~S., Hoyle F., 2001, MNRAS, 325, 483  
%
\bibitem[\protect\citename{Croom2}2002]{croom02}
Croom S.~M., Boyle B.~J., Loaring N.~S., Miller L., Outram P.~J., Shanks T., 
Smith R.~J., 2002, MNRAS, 335, 459 
%
\bibitem[\protect\citename{cral}2002]{cral}
Croom S.~M., Smith R.~J., Boyle B.~J., Shanks T., Miller L., Outram P.~J., Loaring N.~S., 2004, MNRAS, 349, 1397
%
\bibitem[\protect\citename{dad}2001]{dad}
Daddi E., Broadhurst T., Zamorani G., Cimatti A., R\"ottgering H., Renzini, A.,
2001, A\&A, 376, 825
%
\bibitem[\protect\citename{dim}2003]{dim}
Di Matteo T., Croft R.~A.~C., Springel V., Hernquist L., 2003, ApJ, 593, 56
%
\bibitem[\protect\citename{dun}2003]{dun}
Dunlop J.~S., McLure R.~J., Kukula M.~J., Baum S.~A., O'Dea C.~P., Hughes D.~H., 2003, MNRAS, 340, 1095
%
\bibitem[\protect\citename{efst}1998]{efst}
Efstathiou G., 1988, in Lawrence A., eds., proc. 3rd IRAS conf., 
Comets to Cosmology, Springer, New York, p. 312
%
\bibitem[\protect\citename{efre}1988]{efre}
Efstathiou G., Rees M.~J., 1988, MNRAS, 230, 5P
%
\bibitem[\protect\citename{erz}2002]{erz}
Elvis M., Risaliti G., Zamorani G., 2002, ApJ, 565, L75
%
\bibitem[\protect\citename{elv}1994]{elv}
Elvis M.~W., Wilkes B.~J., McDowell J.~C., Green R.~F., Bechtold J., Willner S.~P., Oey M.~S., Polomski E., Cutri R., 1994, ApJS, 95, 1
%
\bibitem[\protect\citename{eno}2003]{eno}
Enoki M., Nagashima M., Gouda N., 2003, PASJ, 55,133
%
\bibitem[\protect\citename{fabiw}1999]{fabiw}
Fabian A.~C., Iwasawa K., 1999, MNRAS, 303, L34
%
\bibitem[\protect\citename{fal}2004]{fal}
Falomo R., Kotilainen J.~K., Pagani C.,Scarpa R., Treves A., 
2004, ApJ, 604, 495
%
\bibitem[\protect\citename{fan1}2001]{fan1}
Fan X. et al. 2001, AJ, 121, 54
%
\bibitem[\protect\citename{fan4}2004]{fan4}
Fan X. et al. 2004, AJ, in press, astro-ph/0405138
%
\bibitem[\protect\citename{ferra}2002]{ferra}
Ferrarese L., 2002, ApJ, 578, 90
%
\bibitem[\protect\citename{fermer}2000]{fermer}
Ferrarese L., Merritt D., 2000, ApJ, 539, L9
%
\bibitem[\protect\citename{firth}2002]{firth}
Firth A.~E et al., 2002, MNRAS, 332, 617
%
\bibitem[\protect\citename{geb}2000]{geb}
Gebhardt K. et al., 2000, ApJ, 539, L13
%
\bibitem[\protect\citename{gra}2004]{gra}
Granato G.~L., De Zotti G., Silva L., Bressan A., Danese L., 
2004, ApJ, 600, 580
%
\bibitem[\protect\citename{GRA}2004]{GRA}
Grazian A., Negrello M., Moscardini L., Cristiani S., Haehnelt M.~G.,
Matarrese S., Omizzolo A., Vanzella E., 2004, AJ, 127, 592
%
\bibitem[\protect\citename{HK}2000]{HK}
Haehnelt M.~G., Kauffmann G., 2000, MNRAS, 318, 35
%
\bibitem[\protect\citename{HNR}1998]{HNR}
Haehnelt M.~G., Natarajan P., Rees M.~J., 1998, MNRAS, 300, 817
%
\bibitem[\protect\citename{HH}2001]{HH}
Haiman Z., Hui L., 2001, ApJ, 547, 27
%
\bibitem[\protect\citename{Ham04}2004]{Ham04}
Hamana T., Ouchi M., Shimasaku K., Kayo I., Suto Y., 2004, MNRAS, 347, 813
%
\bibitem[\protect\citename{Ham93}1993]{Ham93}
Hamilton A.~J.~S., 1993, ApJ, 417, 19
%
\bibitem[\protect\citename{HCT}2003]{HCT}
Hamilton T.~S., Casertano S., Turnshek D.~A., 2002, ApJ, 576, 61
%
\bibitem[\protect\citename{Hatton}2001]{Hatton}
Hatton S., Devriendt J.~E.~G., Ninin S., Bouchet F.~R., Guiderdoni B., 
Vibert D., 2003, MNRAS, 343, 75
%
\bibitem[\protect\citename{Heap}2000]{Heap}
Heap S.~R., Williger G.~M., Smette A., Hubeny I., Sahu M.~S., Jenkins E.~B.,
Tripp T.~M., Winkler J.~N, 2000, ApJ, 534, 69
%
\bibitem[\protect\citename{Hog}1997]{Hog}
Hogan C.~J., Anderson S.~F., Rugers M.~H., 1997, AJ, 113, 1495
%
\bibitem[\protect\citename{Hoy}2002]{Hoy}
Hoyle F., Outram P.~J., Shanks T., Croom S.~M., Boyle B.~J., Loaring N.~S., 
Miller L., Smith R.~J., 2002, MNRAS, 329, 336
%
\bibitem[\protect\citename{Hut}2002]{Hut}
Hutchings J.~B., Frenette D., Hanisch R., Mo J., Dumont P.~J., Redding D.~C., 
Neff S.~G., 2002, AJ, 123, 2936
%
\bibitem[\protect\citename{IS}1988]{IS}
Iovino A., Shaver P.~A., 1988, ApJ, 330, L13
%
\bibitem[\protect\citename{Jak}2003]{Jak}
Jakobsen P., Jansen R.~A., Wagner S., Reimers D., 2003, A\&A, 397, 891
%
\bibitem[\protect\citename{Kai}1987]{Kai}
Kaiser N., 1987, MNRAS, 227, 1 
%
\bibitem[\protect\citename{KH0}2000]{KH0}
Kauffmann G., Haehnelt M.~G., 2000, MNRAS, 311, 576
%
\bibitem[\protect\citename{KH2}2002]{KH2}
Kauffmann G., Haehnelt M.~G., 2002, MNRAS, 332, 529
%
\bibitem[\protect\citename{Ksdss}2003]{Ksdss}
Kauffmann G. et al., 2003, MNRAS, 346, 1055
%
\bibitem[\protect\citename{Kuk}2001]{Kuk}
Kukula M.~J., Dunlop J.~S., McLure R.~J., Miller L., Percival W.~J., 
Baum S.~A., O'Dea C.~P., 2001, MNRAS, 326, 1533 
%
\bibitem[\protect\citename{LC}1993]{LC}
Lacey C., Cole S., 1993, MNRAS, 262, 627
%
\bibitem[\protect\citename{Laf}1998]{Laf}
La Franca F., Andreani P., Cristiani S., 1998, ApJ, 497, 529
%
\bibitem[\protect\citename{LS93}1993]{LS93}
Landy S.~D., Szalay A.~S., 1993, ApJ, 412, 64
%
\bibitem[\protect\citename{Lef}1996]{Lef}
Le F\`evre O., Hudon D., Lilly S.~J., Crampton D., Hammer F., Tresse L.,
1996, ApJ, 461, 534
%
\bibitem[\protect\citename{Lyn}1969]{Lyn}
Lynden-Bell D., 1969, Nature, 223, 690
%
%
\bibitem[\protect\citename{mm}1999]{mm}
Magliocchetti M., Maddox S., 1999, MNRAS, 306, 988
%
\bibitem[\protect\citename{maglio}2003]{maglio}
Magliocchetti M., Porciani C., 2003, MNRAS, 346, 186
%
\bibitem[\protect\citename{maglio2}2004]{maglio2}
Magliocchetti et al., 2004, MNRAS, in press, astro-ph/0312160 
%
\bibitem[\protect\citename{magorr}1998]{magorr}
Magorrian J. et al., 1998, AJ, 115, 2285 
%
\bibitem[\protect\citename{marco}2004]{marco}
Marconi A., Risaliti G., Gilli R., Hunt L.~K., Maiolino R., Salvati M., 2004,
MNRAS, in press, astro-ph/0311619
%
\bibitem[\protect\citename{marinoni} 2001]{marinoni}
Marinoni C., Hudson M.~J., 2002, ApJ,  569, 101
%
\bibitem[\protect\citename{mw}2001]{mw}
Martini P., Weinberg D.~H., 2001, ApJ, 547, 12
%
\bibitem[\protect\citename{Matarrese}1997]{Matarr}
Matarrese S., Coles P., Lucchin F., Moscardini L., 1997, MNRAS, 286, 115
%
\bibitem[\protect\citename{McD}2004]{McD}
McLure R.~J., Dunlop J.~S., 2004, MNRAS, submitted, astro-ph/0310880
%
\bibitem[\protect\citename{Men}2003]{Men}
Menci N., Cavaliere A., Fontana A., Giallongo E., Poli F., Vittorini V.,
2003, ApJ, 587, 63
%
\bibitem[\protect\citename{MF}1993]{MF}
Mo H.~J., Fang L.~Z., 1993, ApJ, 410, 493
%
\bibitem[\protect\citename{Mo}1996]{Mo}
Mo H.~J., White S.~D.~M., 1996, MNRAS, 282, 347
%
\bibitem[\protect\citename{MMW}1998]{MMW}
Mo H.~J., Mao S., White S.~D.~M., 1998, MNRAS, 295, 319
%
\bibitem[\protect\citename{MSD}2000]{MSD}
Monaco P., Salucci P., Danese L., 2000, MNRAS, 311, 279
%
\bibitem[\protect\citename{Moscardini}2000]{Moscard}
Moscardini L., Matarrese S., Lucchin F., Rosati P., 2000, MNRAS, 316, 283
%
\bibitem[\protect\citename{Mou}2002]{Mou}
Moustakas L., Somerville R.~S., 2002, ApJ, 577, 1
%
\bibitem[\protect\citename{NFW}1997]{NFW}
Navarro J.F., Frenk C.S., White S.D.M., 1997, ApJ, 490, 493
%
%
\bibitem[\protect\citename{Norb-b}2002]{Norb-b}
Norberg P. et al. 2002, MNRAS, 336, 907
%
\bibitem[\protect\citename{Om}2001]{Om}
Omont A., Cox P., Bertoldi F., McMahon R.~G., Carilli C., Isaak K.~G.,
2001, A\&A, 374, 371
%
\bibitem[\protect\citename{oual}2003]{oual}
Outram P.~J., Hoyle F., Shanks T., Croom S.~M., Boyle B.~J., Miller L., Smith R.~J., Myers A.~D., 2003, MNRAS, 342, 483
%
\bibitem[\protect\citename{oual2}2004]{oual2}
Outram P.~J., Shanks T., Boyle B.~J., Croom S.~M., Hoyle F., Loaring N.~S., 
Miller L., Smith R.~J., 2004, MNRAS, 348, 745
%
\bibitem[\protect\citename{Peacock} 1996]{peacock}
Peacock J.~A., Dodds S.~J., 1996, MNRAS, 267, 1020
%
\bibitem[\protect\citename{Peacock3} 2000]{peacock3}
Peacock J.~A., Smith R.~E., 2000, MNRAS, 318, 1144
%
\bibitem[\protect\citename{pm} 2001]{pm}
Percival W.~J., Miller L., McLure R.~J., Dunlop J.~S., 2001, MNRAS, 322, 843
%
\bibitem[\protect\citename{Perci} 2003]{perci}
Percival W.~J., Scott D., Peacock J.~A., Dunlop J.~S., 2003, MNRAS. 338, L31  
%
\bibitem[\protect\citename{Porci} 2002]{porci}
Porciani C., Giavalisco M., 2002, ApJ, 565, 24 
%
\bibitem[\protect\citename{Porcietal} 1998]{porcietal}
Porciani C., Matarrese S., Lucchin F., Catelan P., 1998, MNRAS, 298, 1097
%
\bibitem[\protect\citename{rich} 1998]{rich}
Richstone D. et al., 1998, Nature, 395, A14
%
\bibitem[\protect\citename{roch} 2003]{roch}
Roche N.~D., Dunlop J., Almaini O., 2003, MNRAS, 346, 803
%
\bibitem[\protect\citename{salp} 1998]{salp}
Salpeter E.~E., 1964, ApJ, 140, 796 
%
\bibitem[\protect\citename{sal} 1999]{sal}
Salucci P., Szuszkiewicz E., Monaco P., Danese L., 1999, MNRAS, 307, 637 
%
\bibitem[\protect\citename{schir} 2004]{schir}
Schirber M., Miralda-Escud\`e J., McDonald P., 2004, ApJ, 610, 105
%
\bibitem[\protect\citename{sdss} 2003]{sdss}
Schneider D.~P. et al. 2003, AJ, 126, 2579
%
\bibitem[\protect\citename{Scocci}2001]{Scocci}
Scoccimarro R., Sheth R.~K., Hui L., Jain B., 2001, ApJ, 546, 20
%
\bibitem[\protect\citename{sel} 2000]{sel}
Seljak U., 2000, MNRAS, 318, 203
%
\bibitem[\protect\citename{selj}2002]{Selj}
Seljak U., 2002, MNRAS, 334, 797
%
\bibitem[\protect\citename{SB}1994]{SB}
Shanks T., Boyle B.~J., 1994, MNRAS, 271, 753
%
\bibitem[\protect\citename{shal}1987]{shal}
Shanks T., Fong R., Boyle B.~J., Peterson B.~A., 1987, MNRAS, 227, 739
%
\bibitem[\protect\citename{sha}1984]{sha}
Shaver P.~A., 1984, A\&A, 136, L9
%
\bibitem[\protect\citename{sheth }1999]{Sheth}
Sheth R.~K., Tormen G., 1999, MNRAS, 308, 119
%
\bibitem[\protect\citename{sheth1 }2001]{Sheth1}
Sheth R.~K., Diaferio A., 2001, MNRAS, 322, 901
%
\bibitem[\protect\citename{shejai}2003]{SheJai}
Sheth R.~K., Jain B., 2003, MNRAS, 345, 529
%
\bibitem[\protect\citename{sr}1998]{SR}
Silk J., Rees M.~J., 1998, A\&A, 331, L1
%
\bibitem[\protect\citename{Teg}2004]{Teg}
Tegmark M. et al., 2004, PRD in press, astro-ph/0310723
%
\bibitem[\protect\citename{trem}2002]{Trem}
Tremaine S. et al., ApJ, 2002, 574, 740
%
\bibitem[\protect\citename{Vale}2004]{Vale}
Vale A., Ostriker J.~P., 2004, MNRAS, submitted, astro-ph/0402500
%
\bibitem[\protect\citename{vdb}2003]{vdb1}
van den Bosch F.~C., Mo H.~J., Yang X., 2003, MNRAS, 345, 923
%
\bibitem[\protect\citename{Yama}1999]{Yama}
Yamamoto K., Suto Y., 1999, ApJ, 517, 1
%
\bibitem[\protect\citename{ymv}2003]{ymv}
Yang X., Mo H.~J., van den Bosch F.~C., 2003, MNRAS, 339, 1057
%
\bibitem[\protect\citename{YT}2003]{YT}
Yu Q., Tremaine S., 2002, MNRAS, 335, 965
%
\bibitem[\protect\citename{WL3}2003]{WL3}
Wyithe J.~S.~B., Loeb A., 2003, ApJ, 595, 614
%
\bibitem[\protect\citename{WL4}2004]{WL4}
Wyithe J.~S.~B., Loeb A., 2004, ApJ, submitted, astro-ph/0403714
%
\bibitem[\protect\citename{Zehavi} 2003]{zehavi}
Zehavi I. et al., 2003, astro-ph/0301280
%
\bibitem[\protect\citename{ZN}1964]{ZN}
Zel'dovich Ya.~B., Novikov I.~D., 1964, Soviet Phys. Dokl., 158, 811
%
\bibitem[\protect\citename{Zeng}2004]{Zeng}
Zheng Z., 2004, ApJ, in press, astro-ph/0307030
\end{thebibliography}
\end{document}